\documentclass[12pt,twoside]{article}
\usepackage{amssymb,amsmath,amsthm,amscd}
\newcommand{\barefootnote}[1] {%
  \begingroup
    \renewcommand{\thefootnote}{}
    \footnotetext{#1}
    \renewcommand{\thefootnote}{\arabic{footnote}}
  \endgroup
}
\renewcommand{\title}[1] {%
  \begingroup
    \begin{center}
      \vspace{0.4in}
      \bf\huge
      \addtolength{\baselineskip}{5mm}
      #1
    \end{center}
  \endgroup
}

\newcommand{\url}[1] {%
  \barefootnote{%
    {\small e-print archive: }
    {\texttt http://xxx.lanl.gov/#1}
  }
}

\renewcommand{\author}[1] {%
  \begingroup
    \begin{center}
      \vspace{0.4in}
      \bf
      #1
      \vspace{0.2in}
    \end{center}
  \endgroup
}

\newcommand{\address}[1] {%
  \begingroup
    \begin{center}
      #1
    \end{center}
  \endgroup
}
\newcommand{\addressemail}[1] {%
  \begingroup
    \begin{center}
      \vskip-\baselineskip
      #1
    \end{center}
  \endgroup
}
\newcommand{\Gat}{\widetilde{G}}
\newcommand{\gat}{\widetilde{{\mathcal G}}}
\newcommand{\Dinfty}{\mbox{\rm Diff}^{\infty}_{+}(S^{1})}
\newcommand{\Dteps}{\mbox{\rm Diff}^{2+\epsilon}_{+}(S^{1})/%
	\mbox{\rm PSU}_{1,1}}
\newcommand{\Dtp}{\mbox{\rm Diff}^{3/2}_{+}(S^{1})/\mbox{\rm PSU}_{1,1}}

\newcommand{\Di}{{\nEW D}}

\newcommand{\pst}{\widetilde{\psi}}
\newcommand{\cht}{\widetilde{\chi}}

\newcommand{\nEW}[1]{\mathbb{#1}}

\newcommand{\pdx}{\partial_{x}}

\newcommand{\Rin}{{\mathbb R}}
\newcommand{\PR}{{\rm pr}}
\newcommand{\Tr}{{\rm Tr}}

\newcommand{\bydef}{\stackrel{\mbox{\rm \scriptsize def}}{=}}

\newcommand{\dbar}{\bar{\partial}}
\newcommand{\dbphi}{\bar{\partial}_{\phi}}
\newcommand{\ngi}{-\frac{1}{2}{\mathcal S}[f_{\infty}(1/z)]}
\newcommand{\muphi}{-\frac{1}{2}(1 - |z|^{2})^{2}%
{\mathcal S}[f_{\infty}(1/z)](\bar{z})}
\newcommand{\gig}{\gamma_{\psi}}
\newcommand{\HSob}{{\mathcal H}_{{\mathrm{per}}} }
\def\<{{\langle}}
\def\>{{\rangle}}

\begin{document}

\title{Geodesic Flows on Diffeomorphisms of the Circle,
Grassmannians, and the Geometry of the Periodic KdV Equation}

\vspace*{-8mm}
\author{Maria E. Schonbek, Andrey N. Todorov, and\\ Jorge P. Zubelli} 

\vspace*{-5mm} 
\address{UCSC, California, USA}  
\addressemail{schonbek@cats.ucsc.edu} 
\address{UCSC and Bulgarian Academy of Sciences, Sofia}
\addressemail{todorov@cats.ucsc.edu} 
\address{IMPA, Rio de Janeiro, Brazil}
\addressemail{zubelli@impa.br}


\newtheorem{thr}{Theorem} 
\newtheorem{pr}[thr]{Proposition}
\newtheorem{re}[thr]{Remark}
\newtheorem{co}[thr]{Corollary}
\newtheorem{lm}[thr]{Lemma}
\newtheorem{defi}[thr]{Definition}
\newcounter{bean}
\newtheorem{hyp}[bean]{Assumption}

\begin{abstract}
We start by constructing a Hilbert manifold ${\mathcal T}$ of orientation
preserving diffeomorphisms of the circle (modulo the group of bi-holomorphic
self-mappings of the disc).
This space, which  could be thought of as a completion of the universal
Teichm\"uller space, is endowed with a right-invariant K\"{a}hler
metric.
\newpage

Using results from the theory of quasiconformal mappings we construct an
embedding of ${\mathcal T}$ into the infinite dimensional Segal-Wilson
Grassmannian. The latter turns out to be a very natural ambient space for
${\mathcal T}$. This allows us to prove that ${\mathcal T}$'s
sectional curvature is negative in the holomorphic directions and by a 
reasoning along the lines of Cartan-Hadamard's theory that its geodesics
exist for all time.

The geodesics of ${\mathcal T}$ lead to solutions of the periodic
Korteweg-de Vries (KdV) equation by means of V.~Arnold's generalization of
Euler's equation. As an application, we obtain long-time existence of
solutions to the periodic  KdV equation with initial data
in the periodic Sobolev space 
$\HSob^{3/2}(\Rin,\Rin)$.
\end{abstract}

\pagestyle{myheadings}
\markboth{\hfil{\sc \small M.E. SCHONBEK/A.N. TODOROV/J.P. ZUBELLI\quad}\hfil}%
{\hfil{\sc \small \qquad GEODESIC FLOWS on DIFFEOMORPHISMS}\hfil}
\setcounter{page}{2}

\section{Introduction}
\label{sec1}

The interplay between the theory of infinite dimensional 
completely integrable systems and complex analysis has been
extremely fruitful to both fields. 
We feel, however, that there are still many pieces of the picture missing, 
especially if one takes into account the development of the 
theory of Teichm\"uller spaces and quasiconformal mappings. 

Our main object of study concerns geometric aspects of 
a certain class of orientation preserving diffeomorphisms of the circle 
modulo conformal diffeomorphisms of the disk.
More precisely, this Teichm\"{u}ller space ${\mathcal T}$ is endowed with a
Hilbert manifold  structure by means of a (unique up to constant)
right-invariant K\"{a}hler metric and contains the Teichm\"{u}ller space 
of $C^{\infty}$ diffeomorphisms of the circle. The construction of such
space ${\mathcal T}$ is the object of Section~\ref{sec2}.

One of the motivations for such study is the remarkable fact that 
investigating the geodesics of ${\mathcal T}$ will give information about the
behavior of solutions to the periodic Korteweg-de Vries equation
\begin{equation}
\partial_t u = \partial_{x}^{3}u + 6 u \partial_{x} u.
\label{eq:kdv11}
\end{equation}
The connection between the geodesic flow on ${\mathcal T}$ and 
the KdV equation follows from V.~Arnold's approach to
Euler's equation \cite{arnolfou,arnold}, which was applied by
Khesin and Ovsienko~\cite{ovs} to one of the co-adjoint orbits of the
Bott-Virasoro group.  
We apply the same ideas to {\em a different} co-adoint orbit. 
As far as we know, this is a new remark (see also \cite{segal}).
The upshot is the following result, which is proved in Appendix~1:

\begin{thr}
If  ${\mathcal T}$ is endowed with
the above mentioned right-invariant K\"ahler metric,  then one can associate
to each geodesic through the identity 
a solution to the KdV equation. Furthermore, in this case, the solution
to the KdV equation exists for all time.
\end{thr}

To prove this, we shall need some preparation. 
Our first result, which seems to be proved in the
literature \cite{segal} only for the case of $C^{\infty}$
diffeomorphisms, is the following: 

\begin{thr}
The space ${\mathcal T}$ embeds isometrically into the Segal-Wilson
Grassmannian, when the latter is endowed with the Hilbert-Schmidt
norm.
\end{thr}

By virtue of this embedding, we are in position of 
studying the curvature properties of ${\mathcal T}$ in
a very convenient ambient space. This allows us to
prove the following basic result, which is of interest on its own:

\begin{thr}
The sectional curvature of ${\mathcal T}$ in the holomorphic directions is 
bounded from above by a negative number.
\end{thr}

As a corollary, we can identify the Teichm\"uller space ${\mathcal T}$
with the infinite-dimensional Siegel disc.  See Section~\ref{sec6}.

Kirillov and Yurev~\cite{kirillov1, kirillov2} have formulas for 
the curvature of 
${\mathcal T}$, but we believe that this is the first time one shows
the negativity of the sectional curvature in the holomorphic direction
 for the invariant K\"ahler metric of ${\mathcal T}$.

A consequence of this circle of ideas is that the 
existence of geodesics on ${\mathcal T}$ implies 
existence of periodic solutions to the KdV equation.
More precisely, through an application of the Hopf-Rinow theorem
\cite{docarmo,helgason} to a certain two dimensional totally geodesic
complete manifold we obtain long time existence of periodic solutions
to the KdV equation with initial (real) data in 
the $2\pi$-periodic Sobolev space $\HSob^{3/2}(\Rin,\Rin)$. 

This approach has a very different flavor from the standard proof of existence
of solutions to the KdV equation, which can be found for example in 
 \cite{iolisc,kato,kpv1,kpv2,ponce} and references therein. 
We feel that this gives  a very promising geometrical picture of the problem. 
The hope being that this can be applied  to more general problems. 

A good part of the paper is dedicated to the computation of the 
curvature tensor of ${\mathcal T}$.  
We show that sectional curvature 
in the holomorphic directions is less than $-3/2$.
This information is then used together with an extension to 
infinite dimensions of the Cartan-Hadamard theorem to show 
the Arnold exponential instability of the geodesic flow.
The general approach for infinite dimensional case of 
the Cartan-Hadamard theory can be found in S.~Lang's book \cite{langbook}.

This paper will be organized as follows:

In Sections \ref{sec2} and \ref{sec3} we construct a holomorphic
equivariant map from the Teichm\"{u}ller space
\[ {\mathcal T} = \Dtp \] 
to the Segal-Wilson Grassmannian. Here, and 
throughout this paper, we denote by $\mbox{\rm PSU}_{1,1}$
the sub-group of linear fractional transformations that send
the unit disk into itself. The embedding is done by
recalling that each $\phi\in {\mathcal T}$ can be expressed as the
composition of two univalent functions $f_{0}^{-1}$ and $f_\infty$. 
The function $f_{\infty}$ is defined in the exterior
of the disc, which we call  $\Di_{\infty}$, and $f_{0}^{-1}$ is defined on the 
complement of $f_{\infty}(\Di_{\infty})$.
From the function $f_\infty$ we define the Beltrami 
operator. Hence, we  associate to each $\phi$ the space of solutions $W_{\phi}$
to the Beltrami equation with a certain complex dilation $\mu_{\phi}$.\footnote{The idea of looking at 
the elements of the Grassmannian as the
boundary values of solutions of differential equations is due
to Witten \cite{witten1}.}
The functions in $W_{\phi}$ when
restricted to $S^{1}$ are naturally elements of  the Hilbert 
space $H\bydef L^{2}(S^{1})$. Thus, $W_{\phi}$ can be identified
with a subspace of $H$, which we denote by ${\mathcal W}_{\phi}$. 
However, more can be said.
Recall the standard decomposition $H=H_{+}\oplus H_{-}$ as the direct sum of 
non-negative indexed Fourier components and negative indexed Fourier 
components. We show in Section~\ref{sec3}  that the 
projection $\PR_{-}:{\mathcal W}_{\phi}\rightarrow H_{-}$ is Hilbert-Schmidt.
Since from our construction it follows immediately that
$\PR_{+}:{\mathcal W}_{\phi}\rightarrow H_{+}$ is an isomorphism we get that
${\mathcal W}_{\phi}$ is a point in the Segal-Wilson 
Grassmannian \cite{SegalW}.
This Grassmannian, which plays an important role in theory of
solitons, is used here in a totally novel way (for the traditional
approach see \cite{JimboMiwa,SegalW}). In our construction, the 
KdV flow appears as a geodesic flow in ${\mathcal T}$, which can 
be in turn embedded into the Segal-Wilson Grassmannian. 

Section~\ref{sec4} is dedicated to endowing
the space ${\mathcal T}$ with a Hilbert manifold structure.

In Section~\ref{sec5} we show that the Teichm\"{u}ller space ${\mathcal T}$ 
equipped with the unique (modulo a constant) right-invariant K\"{a}hler 
metric has negative Gaussian
curvature. The proof is based on the construction of the so-called
Cartan coordinate system. 

Section~\ref{sec6} is concerned with the 
description of the geodesics. Here we show that since the geodesics
stay in a certain two-dimensional manifold, we can employ the
Hopf-Rinow theorem to establish global existence in time of the geodesics.
Hence, using the results reviewed in  the Appendix~1 we are able to show global
existence of solutions to the periodic KdV equation with initial
data in $\HSob^{3/2}$. Section~\ref{sec6} ends with an infinite dimensional
analogue of Cartan-Hadamard's theorem \cite{docarmo,helgason} showing 
Arnold's exponential spreading of the geodesics.

In Section~\ref{sec7} we give a procedure to construct periodic solutions
to the KdV equation. This is done explicitly, modulo the Riemann mapping
theorem, by using solutions of Beltrami's equation. 

The paper ends with two appendices. On the first one, we give a review of 
Arnold's point of view on Euler's equation and the construction of geodesics.
On the second one, 
we summarize the different results from Teichm\"uller theory used throughout 
the paper.

We  close the introduction with a bit of notation. 
We shall denote by $\HSob^{s}({\nEW R};{\nEW C})$ 
 the set of ($2\pi$-periodic) distributions 
$\sum_{n\in {\nEW Z}} a_{n} \exp(i n x)$ such that 
$\sum_{n\in {\nEW Z}} (1+n^{2})^{s}  |a_{n}|^{2} \le \infty $.
We shall often identify this space with
${\mathcal H}^{s}(S^{1})$ by setting $z=\exp(ix)$.

\section{Construction of a Holomorphic map\\ from 
$\Dtp$ to the\\ Segal-Wilson Grassmannian}
\label{sec2}

We recall that the Segal-Wilson Grassmannian is the set 
of closed subspaces $W$ of $H=H_{+}\oplus H_{-}$  such that the projection
$\PR_{+} : W \rightarrow H_{+}$ is Fredholm and the projection
$\PR_{-} : W \rightarrow H_{-}$ is Hilbert-Schmidt, where H is the
$L^2$ space of complex functions on $S^{1}$. 

The goal of Sections~2 through 5 is to construct the space~\footnote{
The upper index $3/2$ in the above formula is to underscore
that the metric on the tangent space is equivalent to a Sobolev $3/2$ norm.}
\[ {\mathcal T} = \Dtp \mbox{ }  \]
and simultaneously to  embed it into the Segal-Wilson Grassmannian.
This embedding will prove to be instrumental in the
computation of the curvature of ${\mathcal T}$ and the construction
of the Cartan coordinates.
One may think of ${\mathcal T}$ as a  completion of the
space of  $\Dinfty/\mbox{\rm PSU}_{1,1}$ with respect to a
right-invariant K\"ahler metric.  This abstract definition, however,
does not seem to make the embedding into the Grassmannian evident.
We chose therefore to proceed in a more concrete way by constructing
${\mathcal T}$ and its embedding simultaneously.

This approach also highlights different but equally important aspects 
of ${\mathcal T}$. From one side one can think of
${\mathcal T}$ as a space of Beltrami differentials,
from a second one, as an orbit space of the Virasoro group, and from a third 
one as a complete K\"ahler manifold isometrically embedded into the
Segal-Wilson Grassmannian.

In the present section we limit ourselves to  constructing 
a map
 \[ \Psi:  {\mathcal T}\cap {\mathcal D} \rightarrow \widetilde{Gr}, \]
where
\[ \widetilde{Gr} = \{ W \text{ closed subspace of } 
H | \PR_{+}:W \rightarrow H_{+}\text{ is an isomorphism} \}, \]
and ${\mathcal D}$ is an open set close to the identity map of $S^{1}$.

In the next section we show that if $\phi$ is a diffeomorphism of the
circle satisfying the additional assumptions~1 and 2,  which we make
explicit bellow, then 
the image $W=\Psi(\phi)$ has the  property that $\PR_{-}:W \rightarrow H_{-}$ 
is Hilbert-Schmidt.

The outline of the main steps is essentialy the following: 
From a diffeomorphism $\phi$ we construct the corresponding Beltrami 
differential.
Given the Beltrami differential  we construct an operator $\dbphi$  on the
plane, which agrees with the usual $\dbar$ operator outside the disk.
We then construct a suitable (non-orthonormal) 
basis for $\ker{\dbphi}$ and finally restrict ourselves to the unit
circle to obtain $\Psi(\phi) = \ker{\dbphi}$.  

We will show that the vector space generated by elements in
$\ker{\dbphi}$ when restricted to the circle gives us a point in the
Grassmannian. 

We start with some notation.
The open unit disc centered at the origin will be denoted
by $\Di_{0}$. The complement to the closure of $\Di_{0}$
will be
\[\Di_\infty \bydef \{ z \in \widehat{\nEW C} \   |\  |z| > 1 \},  \]
where $\widehat{\nEW C}$ denotes the Riemann sphere ${\nEW C} \cup
\{\infty \}$. 
The set of compactly supported functions of class $C^{k}$ on ${\nEW C}$ 
will be denoted by $C^{k}_{0}({\nEW C})$.
We denote by $\mbox{\rm PSU}_{1,1}$ the group of bi-holomorphic functions
from the disc $\Di_{0}$  into itself.

Our construction is based on the so called {\em sewing problem}, which also 
plays a role in other problems of complex 
analysis \cite{kirillov1,lehto,lehtovirt}.
Given $\phi$ a quasisymmetric homeomorphism  of the circle, find a pair of
homeomorphisms $f_{0}$ and $f_{\infty}$, such that 
\begin{itemize}
\item[a)]
$ f_{0} :  \overline{\Di}_{0} \longrightarrow  f_{0}(
\overline{\Di}_{0}) \subset  
\widehat{\nEW C}$ and 
$f_{\infty} : \overline{\Di}_{\infty} \longrightarrow f_{\infty}(
\overline{\Di}_{\infty}) \subset  \widehat{\nEW C} $,  
where $f_{0}$ and $f_{\infty}$ are conformal in the interior of their 
domains of definition, 
\item[b)] The sets $f_{0}( \Di_{0})$ and 
$f_{\infty}( \Di_{\infty})$ are complementary Jordan domains,  and
\item[c)] For every $z\in S^{1}$ 
\[\phi(z) = f_{0}^{-1} \circ  f_{\infty}(z) \mbox{ ,} \]
where the superindex $-1$ indicates the inverse function.
\end{itemize}
In the Appendix~2 we shall review the notion of quasisymmetric functions
and cover some known results related to the theory of Teichm\"uller spaces.
For the time being we remark that the sewing problem stated above has
a unique solution provided we require $\phi$, $f_{0}$ and $f_{\infty}$
to be normalized. One such normalization is achieved by requiring that 
they all fix the points
$-1$, $-i$ and $1$ of $S^{1}$. See \cite{pfluger,lehto,lehtovirt}.

We recall that the Schwarzian derivative of an analytic function $f$ is 
defined by 
\[ {\mathcal S}(f) \bydef \frac{f'''}{f'} -
\frac{3}{2}\left(\frac{f''}{f'}\right)^{2} .\] 

We are now ready to introduce a key object in our construction:

\begin{defi}\label{def1}
{\em
For $\phi$ a normalized quasisymmetric function we set 
\[ 
\mu_{\phi}(z) =
\begin{cases}

\muphi,& \quad z \in \Di_{0} \\
0, & \quad z \in {\nEW C} \setminus \Di_{0}. 
\end{cases}
\]
}
\end{defi}

Since $f_{\infty}$ is univalent, Nehari's Lemma (see \cite{Ahlfors}) 
implies that 
\[ \| \mu_{\phi} \|_{\infty} \le 3. \]
We assume the following:

\begin{hyp}\label{hypone}
{\em
The function $\phi$ is such that the solution of the sewing 
problem $f_\infty$ satisfies 
\begin{equation}
\label{neha}
\| \mu_{\phi} \|_{\infty} < 1 .
\end{equation}
}
\end{hyp}

In Appendix~2 we show that if $\phi$ is sufficiently close
to the identity map in the $C^{1}$ topology, then Assumption~\ref{hypone}
is satisfied.

We define the Beltrami operator:
 
\begin{defi} 
{\em
Let $\mu \in L^{\infty}({\nEW C})$. For 
$f \in C^{\infty}_{0}({\nEW C})$ we define the operator 
$\dbar_{\mu}$ by 
\[ \dbar_{\mu} f (z)  \bydef  \dbar f (z) - \mu(z)  \partial f(z). \] 
}
\end{defi}

We shall denote also by $\dbar_{\mu}$ the extension (in the
weak-derivative sense) 
of this operator to more general functions, such as $f \in L^{2}({\nEW C})$ 
whose distributional derivatives $\partial f$ and $\dbar f$ are locally in 
$L^{2}$. 

Recall that $H_{+}$ denotes 
\[ H_{+}=\left\{ f\in L^2(S^{1}) | f(z)= \sum_{n\ge 0} a_{n}z^{n},
 z\in S^{1}  \right\} , \] 
and $H_{-}$ denotes its orthogonal complement in $L^{2}(S^{1})$.
Hence,
$ H=H_{+}\oplus H_{-} .$

\begin{thr} 
The map $\phi \rightarrow {\ker{\dbar}}_{\mu(\phi)}|_{S^1}= W_{\phi}
\subset H$ defines an isometric embedding of  $\Dinfty/\mbox{\rm
PSU}_{1,1}$ into the Segal-Wilson Grassmannian endowed with the
Hilbert-Schmidt norm. (The metric on 
$\Dinfty/\mbox{\rm PSU}_{1,1}$ will be defined in Section 4.)
\end{thr}

\noindent 
The main step in the proof of this theorem, which will be given in 
Section 3, is to show that $pr_{-} : W_{\phi} \rightarrow H_{-}$ is
Hilbert-Schmidt. The proof of this fact  relies heavily on the
construction of a special basis for W. The rest of this section is
devoted to this construction. 

We define some operators that will play an important role.

\begin{defi}
{\em
Let $h\in L^{p}({\nEW C})$, then
\begin{equation}  \label{defP}
P h (\zeta)  \bydef -\frac{1}{\pi}\int_{\nEW R} 
\int _{\nEW R} h(z)\left(\frac{1}{z-\zeta}\right)\; dx dy ,
\end{equation}
where $z=x+iy$.
}
\end{defi}

\begin{defi}
{\em
Let $h\in C_{0}^{2}({\nEW C})$, then 
\[ T h (\zeta)  \bydef \lim_{\epsilon\rightarrow 0}
-\frac{1}{\pi} \int \int _{|z-\zeta|^{2}>\epsilon}
\frac{h(z)}{(z-\zeta)^{2}} dx dy ,  \]
where $z=x+iy$.
}
\end{defi}

It can also be shown that (Lemma~2 page 87 of \cite{Ahlfors}) 
\[ \dbar P h = h , \]
and
\[ \partial P h = T h \mbox{.} \]
From a result of Calderon and Zygmund it is known 
that $T$ extends as a bounded operator from $L^{p}({\nEW C})$
into itself of norm $C_{p}$ for any $p>1$, where $C_{p}\rightarrow 1$
as $p\rightarrow 2$.

\medskip\noindent
{\it Construction of Special Solutions to Beltrami's Equation:} 

We define $\nu^{(n)}$ by
\[ \nu^{(n)} = \sum_{k=1}^{\infty} T^n_{k}(\mu_{\phi}) \mbox{,}\] 
where
$T_{0}^{n}(\mu_{\phi})=nz^{n-1}$ and
\[ T^{n}_{k}(\mu_{\phi}) = T(\mu_{\phi}(T^{n}_{k-1}(\mu_{\phi}))) \]
Now we set, for $n> 0$, 
\begin{equation}
w^{(n)} \bydef 
z^{n}+P(\mu_{\phi}(\nu^{(n)}+nz^{n-1})) \mbox{.}
\label{eqqeq1}
\end{equation}

The following result is proved for the case of $n=1$ in
\cite{Ahlfors} (Theorem 1, page~91). The general case is a 
generalization of the proof therein and will be 
given in the Appendix~B.

\begin{thr}
For any integer $n\ge 1$ we have that $w^{(n)}$ is the unique solution
of the problem:
\begin{equation}
\label{eqdbar}
 \left\{ \begin{array}{l}
\dbar_{\mu_{\phi}} w^{(n)}  = 0  \mbox{,} \\
\partial_{z} w^{(n)} - n z^{n-1} \in L^{p} \mbox{ for some } p >  2,  \\
\int_{0}^{2\pi} w^{(n)}(e^{i \theta})d\theta  = 0.
\end{array}
\right. \end{equation}
\end{thr}

Note that from equation (\ref{eqqeq1})
\[ \mu_{\phi}(\nu^{(n)}+nz^{n-1}) = 
\mu_{\phi}  \sum_{k=0}^{\infty}
(T\circ \mu_{\phi})^{k} ( n z^{n-1}) \mbox{,}\]
where, by $T\circ \mu_{\phi}$ we mean the composition
of the operator $T$ with the operator of multiplication by $\mu_{\phi}$.
So, with this notation, we get another 
way of writing $w^{(n)}$.
It can be written as
\begin{equation} 
\label{smart}
w^{(n)}=z^{n}+P\left[\mu_{\phi}  \sum_{k=0}^{\infty} 
(T\circ \mu_{\phi})^{k} ( n z^{n-1}) \right]  \mbox{.} \end{equation} 

\begin{lm}\label{comp}
The restriction to $S^{1}$ of second term on the r.h.s. of 
equation~(\ref{smart}) is an element $H_{-}$.
\end{lm}

\begin{proof} 
Since  $w^{(n)}|_{\Di_{\infty}}$ is complex analytic we have that
\begin{equation}
 w^{(n)}\Big|_{\Di_{\infty}} = z^{n} + f(z) + 
 \sum_{j>1}a_{j} z^{-j} \mbox{,} 
\label{eq:eq11}
\end{equation}
where $f(z)$ is an entire function.
From the condition that $ \partial_{z} w^{(n)} - n z^{n-1} \in L^{p}$
it follows that $f'(z)$ is zero. Hence, $f(z)=a_{0}$, where $a_{0}$
is a constant.  From the chosen normalization for the 
definition of the operator $P$ in equation~(\ref{defP}) it is easy to
see that 
\[ a_{0}=\frac{1}{2\pi i} \int_{S^{1}} 
P\left[\mu_{\phi}\left(\nu^{(n)}+n z^{n-1}\right)\right]
(\zeta)\frac{d\zeta}{\zeta}=0 \mbox{.}\]
\end{proof}

\begin{defi}\label{crucial}
{\em
The space ${\mathcal W}_{\phi}$ is defined as the graph of the closure of
the operator
$w_{\phi}:H_{+}\rightarrow H_{-}$ that maps
$z^{n} \mapsto v^{(n)}$, where  for $n =1, 2, \ldots $
\[ v^{(n)}\bydef  P\left.\left[ \mu_{\phi}  \sum_{k=0}^{\infty}
(T\circ \mu_{\phi})^{k} ( n z^{n-1})\right]\right|_{S^{1}}   \]
and $v^{(0)}=0$.
}
\end{defi}

\medskip\noindent
{\bf Remarks.} \  
\begin{enumerate}
\item 
The space ${\mathcal W}_{\phi}$ is spanned by 
the set $w^{(i)}=z^{i}+v^{(i)}$, since it is the graph
of the operator $w_{\phi}$. The main result of the next section
is that $w_{\phi}$ is Hilbert-Schmidt. This implies in particular that
${\mathcal W}_{\phi}$ is closed.
From the definition of ${\mathcal W}_{\phi}$, it follows that
$\PR_{+}:{\mathcal W}_{\phi} \rightarrow H_{+}$ is an isomorphism.
Hence, $\PR_{-}:{\mathcal W}_{\phi}\rightarrow H_{-}$ is a Hilbert-Schmidt 
operator. So the results of the next section show that 
\begin{equation}
 \phi \mapsto {\mathcal W}_{\phi} 
\label{eqmap}
\end{equation}
is a well defined map from ${\mathcal T}$ into the
Grassmannian $Gr$.
\item
In Section~\ref{sec4} we shall show that  the map defined in
equation~(\ref{eqmap}) is holomorphic. Furthermore, we show that
when restricted to $\phi \in C^{\infty}$ it  coincides with 
the map defined in \cite{nag}. From the definition used in 
\cite{nag} it follows  directly the equivariance of this map.
\end{enumerate}

\section{The Projection $\PR_{-}: {\mathcal W}_{\phi} \rightarrow
H_{-}$\\  is Hilbert-Schmidt}
\label{sec3}

The main goal of this section is to prove that under appropriate assumptions
the operator $w_{\phi}$ is a Hilbert-Schmidt operator. 
In Section~\ref{sec4} we prove that all the operators $w_{\phi}$ in a 
certain neighborhood of the identity have such property. 

We recall that for $H_{1}$ and $H_{2}$
Hilbert spaces, an operator $T: H_{1}\rightarrow H_{2}$
is called Hilbert-Schmidt if for one orthogonal basis 
$\{e_{i}\}_{i\in I}$ of $H_{1}$ we have
\[ \sum_{i\in I} \|Te_{i}\|^{2} < \infty \mbox{ .} \] 
If this is the case for one basis of $H_{1}$ it is also the 
case for every basis of $H_{1}$. 

For technical reasons, which will become clear bellow, we
shall need to use the following hypothesis:~\footnote{
In fact, from the proof bellow, such assumption may
be relaxed by asking that $g(z)$ belongs to the (inner) Hardy class
$\mbox{\sf H}^{3/2}(\Di_{0})$, i.e.,
$ \sup_{r < 1} \int_{-\pi}^{\pi} |g(r e^{i\theta})|^{3/2} \ \ 
d \theta < \infty$.
This condition is obviously a consequence of the Assumption~\ref{sechyp}.}

\begin{hyp}
{\em
The function $g=\ngi$ is bounded in $\Di_{0}$.
\label{sechyp}
}
\end{hyp}

We can now state the main result of this section, namely:
 
\begin{thr}
\label{th8}
If $\phi$ satisfies Assumptions~\ref{hypone} and \ref{sechyp}
the operator $w_{\phi}$ of Definition~\ref{crucial}
is a Hilbert-Schmidt operator.
\end{thr}

\medskip\noindent
{\it Proof of the Theorem.} \  
The proof of the theorem is based on showing that under the
hypothesis the sequence $\{v^{(n)}\}_{n\ge 1}$  of 
Definition~\ref{crucial} satisfies 
\[ \sum_{n=1}^{\infty} \|v^{(n)}\|^{2}   < \infty \mbox{.} \]  
Since the space $H_{-}={\mbox span}\{ z^{-i} | i \ge 1\} $, we are going to 
show that
\[ \sum_{n=1}^{\infty}  \sum_{j=1}^{\infty} | \< v^{(n)},z^{-j} \> |^{2} <
\infty \mbox{.} \]
Let's massage a little bit the expression for
$| \< v^{(n)},z^{-j} \> |^{2}$.
We are going to use the convention that $z= \rho \exp( i x)$.

An elementary application of Stoke's theorem gives that 
\begin{eqnarray*}
\< v^{(n)},z^{-j} \> & = &
\frac{1}{2 \pi}\int_{0}^{2 \pi} v^{(n)}(z,\bar{z}) \bar{z}^{-j} \ d x\\
                   & = & 
\frac{1}{2 \pi i}\int_{S^{1}}  v^{(n)}(z,\bar{z}) z^{j} \frac{d z}{z} \\
                   & = &
\frac{1}{2 \pi i}\int_{\Di_{0}} dv^{(n)} \wedge (z^{j-1} dz)  \\
                   & = &
\frac{1}{2 \pi i}\int_{\Di_{0}} \dbar v^{(n)} \ d \bar{z} \wedge 
 (z^{j-1} dz) \mbox{.} 
\end{eqnarray*}

But, from the construction of $v^{(n)}$ we have that
\begin{align*}
\dbar v^{(n)} & = \dbar P \left[\mu_{\phi}
\sum_{k=0}^{\infty} (T\circ \mu_{\phi})^{k}( n z^{n-1})\right]\\ 
& =  \mu_{\phi} \sum_{k=0}^{\infty} 
(T\circ \mu_{\phi})^{k} ( n z^{n-1}) \mbox{,}
\end{align*}
where, by $T\circ \mu_{\phi}$ we mean the composition
of the operator $T$ with the operator of multiplication by $\mu_{\phi}$.
So,
\begin{equation} 
\< v^{(n)},z^{-j} \> = \frac{1}{2 \pi i} 
\int_{\Di_{0}} \mu_{\phi} \sum_{k=0}^{\infty} 
(T\circ \mu_{\phi})^{k} ( n z^{n-1}) z^{j-1} \ d\bar{z} \wedge d z  \mbox{.}
\label{eq:prep}
\end{equation}
We now recall that
\begin{equation}
\mu_{\phi}(z) = (1-|z|^{2})^{2} g(\bar{z}) \mbox{,} 
\label{defmu}
\end{equation}
where 
\begin{equation}
g(z) =  \ngi (z) \mbox{.}
\label{defg}
\end{equation}

Now, we are going to apply Cauchy-Schwarz (for the inner-product of
$L^{2}(\Di_{0})$)
to the  right hand side of  equation (\ref{eq:prep}). Indeed,
we write
$\mu_{\phi}= \mu_{\phi}^{1/4} \mu_{\phi}^{3/4}$, 
except for possibly a set of zero measure in $\Di_{0}$,
and we get 
\begin{align} 
& \left| \frac{1}{2 \pi i} 
\int_{\Di_{0}} \mu_{\phi} \sum_{k=0}^{\infty} 
(T\circ \mu_{\phi})^{k}(n z^{n-1})z^{j-1}d
\bar{z}\wedge d z \right|^2\nonumber\\
& \le \|\mu_{\phi}^{1/4} z^{j-1}\|^{2} 
\left\|  \sum_{k=0}^{\infty} (T\circ \mu_{\phi})^{k} ( n z^{n-1}) 
\mu_{\phi}^{3/4}\right\|^{2} 
\mbox{,} 
\label{eq:prod}
\end{align}
where the norms are taken in $L^{2}(\Di_{0})$.
We are going to estimate each of the norms of equation
(\ref{eq:prod}).
First, $\|\mu_{\phi}^{1/4} z^{j-1}\|^{2}$.
\begin{eqnarray*}
\|\mu_{\phi}^{1/4} z^{j-1}\|^{2} 
& = & \frac{1}{\pi} \int_{0}^{2 \pi} \int_{0}^{1}
[(1-\rho^{2})^{2/4}]^{2} |g(1/\bar{z})|^{2/4} 
\rho^{2j-1}  \   d \rho \ dx \\
& \le &  2 \|g\|_{L^{\infty}(\Di_{0})}^{1/2}
 \int_{0}^{1} (1-\rho^{2})\rho^{2j-1}\ d \rho \\
& =  &   \|g\|_{L^{\infty}(\Di_{0})}^{1/2} \frac{1}{j(j+1)} \\
& =  & {\mathcal O}(1/j^{2}) \mbox{.}
\end{eqnarray*}

Now, we work on the term
$\|  \sum_{k=0}^{\infty} (T\circ \mu_{\phi})^{k} ( n z^{n-1}) 
\mu_{\phi}^{3/4}
\|^{2}$.
We begin by noticing that 
\[ \mu_{\phi}^{3/4} \sum_{k=0}^{\infty} (T\circ \mu_{\phi})^{k} ( n z^{n-1}) 
=  \sum_{k=0}^{\infty} (\mu_{\phi}^{3/4} \circ T \circ \mu_{\phi}^{1/4})^{k} 
(\mu_{\phi}^{3/4} n z^{n-1}). \]
Now, we proceed as in the construction of $w^{(n)}$ (Section~\ref{sec3})  
by using the fact 
that the Hilbert transform $T$ is an isometry of $L^{2}({\nEW C})$, 
to get
\begin{align*}
\left\| \sum_{k=0}^{\infty}(\mu_{\phi}^{3/4}\circ T 
\circ \mu_{\phi}^{1/4})^{k} (\mu_{\phi}^{3/4} n z^{n-1}) \right\| 
& \le \sum_{k=0}^{\infty} \| \mu_{\phi}\|^{k}_{L^{\infty}(\Di_{0})}
\|nz^{n-1}\mu_{\phi}^{3/4}\|.\\
\intertext{So,} 
\left\| \sum_{k=0}^{\infty}(\mu_{\phi}^{3/4}\circ T 
\circ \mu_{\phi}^{1/4})^{k} (\mu_{\phi}^{3/4} n z^{n-1}) \right\|^{2} 
&\le \frac{1}{(1-c_{0})^{2}} \|nz^{n-1}\mu_{\phi}^{3/4}\|^{2}, 
\end{align*}
where
$c_{0} \bydef \|\mu_{\phi}\|_{L^{\infty}(\Di_{0})} < 1$.

Now, we estimate  
\begin{eqnarray*}
 \|nz^{n-1}\mu_{\phi}^{3/4}\|^{2} &  =  & 
\frac{1}{\pi} \int_{0}^{2 \pi} \int_{0}^{1}
n^{2}[(1-\rho^{2})^{3/2}]^{2}|g(\bar{z})|^{3/2}\rho^{2n-1}\,  d \rho \ dx  \\
 & \le & \left(2 n^{2} \int_{0}^{1}(1-\rho^{2})^{3}\rho^{2n-1}\, d\rho\right)
\|g\|_{L^{\infty}(\Di_{0})}^{3/2}\\
& = &  \frac{6n^2}{n(n+1)(n+2)(n+3)} \|g\|_{L^{\infty}(\Di_{0})}^{3/2}\\
& = & {\mathcal O}(1/n^2) \mbox{,}
\end{eqnarray*}
where in the next to last step above we used the fact
that
\[ \int_{0}^{1}(1-\rho^{2})^{3}\rho^{2n-1} \ d\rho  =
\frac{3}{n(n+1)(n+2)(n+3)} \mbox{.} \]

We conclude the proof using the two estimates above
for the norms that appear in the right hand side of 
equation (\ref{eq:prod}) to get
\begin{eqnarray*}
 \sum_{n=1}^{\infty} \|v^{(n)}\|^{2} & \le & 
\sum_{n=1}^{\infty} \sum_{j=1}^{\infty} 
\left( \frac{C_{1}C_{2}}{j^{2}n^{2}} \right) \\
& = &  C_{1}C_{2} \left(\sum_{n\ge1} \frac{1}{n^{2}}\right)^{2}  < \infty
\mbox{,}
\end{eqnarray*}
where $C_{1}$ and $C_{2}$ are constants that depend only on $\phi$.
\endproof

\begin{co}
The projection
$\PR_{-}: {\mathcal W}_{\phi} \rightarrow H_{-}$ is Hilbert-Schmidt.
\end{co}

We remark that the constants $C_{1}$ and $C_{2}$ in the
proof of Theorem~\ref{th8} depended only on the norms
$\|g\|_{L^{\infty}(\Di_{0})}$ and 
$\|\mu_{\phi}\|_{L^{\infty}(\Di_{0})}$
where $g$ and $\mu_{\phi}$  were defined by equations (\ref{defg})
and (\ref{defmu}). A straightforward corollary of this proof
is the following:

\begin{thr}
\label{ext}
Let $g$  be constructed according 
to Section~\ref{sec2}.  Suppose that $g$ is an element of the Hardy class 
$\mbox{\sf H}^{3/2}(\Di_0)$ 
and that \[ \mu_{\phi}(z) = (1-|z|^{2})^{2} g(\bar{z}) \mbox{ }\]
satisfies
 \[ \|\mu_{\phi}\|_{L^{\infty}(\Di_{0})} \le c_{0} < 1 \mbox{.} \]
Then, the operator $w_{\phi}$ of Definition~\ref{crucial} is 
Hilbert-Schmidt and
\[ \|w_{\phi}\|^{2}_{HS} \le 
\frac{K }{ (1-c_{0})^{2} }\|g\|_{ \mbox{\sf H}^{3/2}(\Di_{0}) } 
\|g\|_{ \mbox{\sf H}^{1/2}(\Di_{0}) } \mbox{,} \]
where $K$ is independent of $g$.
\end{thr}

\section{The Hilbert Manifold Structure of\\  $\Dtp$}
\label{sec4}

Our goal in this section is to endow the manifold 
${\mathcal T}\!\!=\Dtp$ with a Hilbert manifold structure.
This will be done in several steps. 

\medskip\noindent
{\bf Step 1.} \ 
Give  a  Hilbert space structure to the
tangent space of the manifold ${\mathcal T}$ at the identity, which
will be denoted by $T_{id} {\mathcal T}$.

Let ${\mathcal T}^{\infty} $ be the space 
of $C^{\infty}$ orientation preserving diffeomorphisms of $S^{1}$ modulo
the group $PSU_{1,1}$.  The space ${\mathcal T}^{\infty} $
has  a complex structure. By that we mean, one has a global 
splitting of the complexified tangent bundle
\[T {\mathcal T}^{\infty} \otimes {\nEW C} = 
(T {\mathcal T}^{\infty}\otimes {\nEW C} )^{(1,0)}
\oplus (T {\mathcal T}^{\infty}\otimes {\nEW C} )^{(0,1)}  \]
such that
\[ \overline{(T {\mathcal T}^{\infty}\otimes {\nEW C} )^{(1,0)}} = 
(T {\mathcal T}^{\infty}\otimes {\nEW C} )^{(0,1)} \]
and that the bracket of two $(1,0)$ vector fields is also $(1,0)$.
  Now, the holomorphic part 
 $ (T_{id} {\mathcal T}^{\infty}\otimes {\nEW C})^{(1,0)} $, 
can be identified by means of  
 \[ (T_{id} {\mathcal T}^{\infty})^{(1,0)} \cong  
\left\{ f \in C^{\infty}(S^{1}; {\nEW C}) | 
 f = \sum_{n\ge 2} a_{n} z^{n} \right\}   \]
To be concrete, we may think of an orientation
preserving  diffeomorphism $\gamma$ on $S^1$
as $\gamma(e^{ix})=\exp(i\psi(x,t))$, where $\psi(x+2\pi)=\psi(x)+2\pi$.
Using such parametrization, a  tangent vector to
${\mathcal T}$ at the origin is given by
\[ \dot{\psi} = \sum_{k\ne \pm 1 , 0} a_{k} e^{ikx} \frac{d}{dx}
\mbox{,} \]
where $a_{k}=\overline{a_{-k}}$ since $\dot{\psi}$ is real.
Then, the corresponding vector in
$(T {\mathcal T}^{\infty}\otimes {\nEW C} )^{(1,0)}$
is given by
$ \sum_{k\ge 2} a_{k} z^{k} $.

We recall that a  Hermitean metric on a complex manifold is called
K\"{a}hler if its imaginary part is a symplectic form.
In \cite{todortudor} it is shown that there exists, up to a constant,
a unique right-invariant K\"{a}hler 
metric on ${\mathcal T}^{\infty} $
such that for any 
$f = \sum_{n\ge 2} a_{n} z^{n} \in (T_{id}{\mathcal T}^{\infty})^{(1,0)} $ 
we have 
\begin{equation}
 \|f\|^2 =  \sum_{n\ge 2} n(n^{2}-1) | a_{n} |^{2} \mbox{.}
\label{kahler}
\end{equation}

We define the Hilbert space $T_{id} {\mathcal T}$
as the completion of the $ T_{id}{\mathcal T}^{\infty} $ with respect
to this metric. It is easy to see that this is 
a linear subspace of $H_{+}$ which is complete in the Sobolev 
$\HSob^{3/2}$-norm. 

\medskip\noindent
{\bf Step 2.} \ 
Define an ``exponential map":

Set 
\[  e_{k}(z) = \frac{z^k}{\sqrt{k(k^{2}-1)}}  \mbox{.}\] 
Obviously, the set $\{e_{k} | k \ge 2\}$ is an o.n. basis of
$ (T_{id} {\mathcal T})^{(1,0)}$ with respect 
to the metric defined in (\ref{kahler}). 
Therefore, any $f$ in such space 
can be written as:
\[ f = \sum_{k\ge 2} t_{k} e_{k} \]
with the vector $(t_{2},t_{3},\dots) \in {\ell}^{2}$. 
We associate to the above $f$ 
\begin{align*}
\mu_{t}(z) & =  \mu(z;t_{2},t_{3},\dots)\\
& \bydef  \begin{cases}
(1-|z|^{2})^{2}  \sum_{k \ge 2} t_{k} e_{k}(\bar{z}) 
\bar{z}^{-2}, & \quad  z\in \Di_{0} \\
 0,  & \quad  z \in \widehat{\nEW C}\setminus \Di_{0}.\end{cases}
\end{align*}
The $\mu_{t}$ constructed this way is a Beltrami differential  provided
that 
\begin{equation}
 \|\mu_{t}(\cdot) \|_{L^{\infty}({\nEW C})} < 1 \mbox{ .} 
\label{eq:condc}
\end{equation}

Note that the set ${\mathcal O}$ of such values of the parameters $t
\bydef (t_{2},t_{3},\dots)$ 
is open in ${\ell}^{2}$.
Indeed, one can easily estimate 
\[ \|\mu_{t}(.) \|_{L^{\infty}} \le  C' \|t\|_{\ell^2} \mbox{.} \]
It is a well known fact that \cite{Ahlfors,bersjohn} for each 
$t\in {\mathcal O}$ there exists a 
quasiconformal diffeomorphism $\omega_{t}$ such that: 
\begin{equation}
\left\{ \begin{array}{l}
\dbar \omega_{t} - \mu_{t}  \partial \omega_{t} = 0  \mbox{,} \\
\omega_{t}(\rho) = \rho \mbox{ for   } \rho \in \{ -1,-i,1\}. 
\end{array}
\right.
\label{eqnorm}
\end{equation}

We remark that this is a bit different from the usual normalization
\[ \left\{ \begin{array}{l} \partial \omega_{t}(z) -1 \in L^{p} 
\mbox{ for some } p >  2  \\
 \omega_{t}(0) = 0 \mbox{,} \end{array}  \right. \]
but the two are related by a linear fractional transformation.

Since $\mu_{t}(z) \equiv 0 $ on $\Di_{\infty}$, the function 
$\omega_{t}$ is  univalent and complex analytic on $\Di_{\infty}$.  
Let 
\[ f_{\infty,t} \bydef  \omega_{t}|_{\overline{\Di}_{\infty}} \mbox{.} \] 
The Riemann Mapping Theorem yields that there 
exists a unique univalent complex analytic function $f_{0,t}$ 
such that
\begin{equation}
\label{rie0}
\left\{ \begin{array}{l}
 f_{0,t}: \Di_{0} \mapsto \widehat{\nEW C} \setminus 
f_{\infty,t}(\Di_{\infty}) \mbox{,} \\
f_{0,t}(\rho) = \rho \mbox{ for   } \rho \in \{ -1,-i,1\}
\end{array}
\right.
\end{equation}

Let
\[ \exp(t) =  f_{0,t}^{-1}  \circ f_{\infty,t} |_{S^{1}} = \sigma_{t}  
\mbox {,} \]
where the super-index $-1$ means the inverse function.

For further use we make note of the following estimates.

\begin{lm}
\label{estim}
If we set, for $z \in \Di_{0}$
\[g(z) = \sum_{k \ge 2} t_{k} \frac{\bar{z}^{k-2}}{\sqrt{k(k^{2}-1)}} 
\mbox{,}\]
then
\[ \|g\|_{L^{\infty}(\Di_{0})} \le C \|t\|_{\ell^{2}} \mbox{.} \]
Also, for $0 \le \epsilon <1$ 
\[ \|\mu_{t}\|_{C^{1+\epsilon}({\nEW C})} 
\le C'_{\epsilon} \|t\|_{\ell^{2}} \mbox{.} \]
\end{lm}

Since the function $\mu_{t}(\cdot)$ defined above is of class
$C^{1+\epsilon}$ for any  $0 < \epsilon < 1$, a straightforward 
adaptation of Theorem~1 on page 269 of \cite{bersjohn} gives that:

\begin{pr}
The quasiconformal mapping $z \mapsto \omega_{t}(z)$ defined \linebreak
above is of class $C^{2+\epsilon}$ for every $0 < \epsilon < 1$. 
\end{pr}
 
As a consequence of this last fact, we can use a result of 
Kellog-Warschawski \cite{war32} (see also page 49 of \cite{pomme})
that ensures that the inner mapping $f_{0,t}$ defined above is
of class $C^{2+\epsilon}$ on $\Di_{0}$.
Hence, our manifold ${\mathcal T}$ will be composed of 
class $C^{2+\epsilon}$ diffeomorphisms. In particular, $\sigma_{t}$
is quasisymmetric.

As an application of the Ahlfors-Weill result (see Appendix~2, 
Theorem~\ref{aw}) it follows that 

\begin{pr}
For $t$ in a neighborhood of $0 \in \ell^{2}$ the mapping 
\[ t\mapsto \exp(t) \in \Dteps \] is locally one-to-one.
\end{pr}

\medskip\noindent
{\bf Step 3.} \ 
Around each point $\phi$ in ${\mathcal T}^{\infty}$ we define a 
new neighborhood $N_{\phi;\epsilon}$ obtained by
right translation of $N_{\sigma_{0};\epsilon}$ by the function $\phi$, where 
$\sigma_{0}$ is the identity map.
In such a way we define the Hilbert manifold ${\mathcal T}$.

Note that  for each $\phi \in {\mathcal T}$, the
tangent space at $\phi$ is given by the span of the following
orthonormal set:
\[ \{   e_{k} \circ \phi| k \ge 2 \}  \mbox {.} \]
Indeed, this can easily be seen by looking at the
tangent space as  equivalence classes of smooth paths in the manifold.

\begin{thr}
The space ${\mathcal T}$ is a complete metric space with respect to
the K\"{a}hler  metric.
\end{thr}

\begin{proof} 
Let $\{\phi_{n}\}$ be any Cauchy sequence.
Using the homogeneity of ${\mathcal T}$ we can assume that
$\phi_{n}$ is contained in the set $N_{0}=\exp(B(0;\epsilon))$,
where $B(0,\epsilon)$ is a sufficiently small ball in 
 ${\ell}^{2}$.
Since the exponential is 1-1 and a local diffeomorphism, it follows that 
the limit of $\exp^{-1}(\phi_{n})$ exists. Hence, $\{\phi_{n}\}$ has a limit.
\end{proof}

\medskip\noindent
{\bf Remarks.} \ 
\begin{enumerate}
\item
The estimate of Lemma~\ref{estim} together with
Theorem~\ref{ext} and the Ahlfors-Weill result gives that
the mapping 
\[{t} \in {\mathcal O} \subset \ell^{2} 
\mapsto {\mathcal W}_{\sigma_{t}} \in \mbox{\rm Gr} \]
is continuous, provided we endow $Gr$ with the K\"ahler metric of 
Section~7.8 of \cite{psegal}.
\item
Along the lines of the Ahlfors-Bers'  theorem \cite{Ahlfors}: If the Beltrami 
differential 
depends holomorphically on the parameter $t$  then the solutions
of equation~(\ref{eqdbar})
also depend holomorphically on $t$. Combining this theorem with the
construction of the exponential map it follows easily that the
map from ${\mathcal T}$ into $Gr$ defined above is holomorphic.

\end{enumerate}

We know	that
\[
\left\{e_{k}=\frac{z^{k}}{\sqrt{k(k^{2}-1)}}\,
\Big|\,k\geq 2\right\}\mbox {.}
\]
is an orthonormal basis	in the tangent space to	$id\in {\mathcal T}.$ In the
next Theorem we	will compute the identification	between	the tangent and
cotangent spaces with respect to the canonical K\"{a}hler metric on 
${\mathcal T} $ which we will call the Weil-Petersson metric.

\begin{thr} \label{t19}
The identification of tangent space to the identity of the Universal
Teichm\"{u}ller  space ${\mathcal T}$ with the	cotangent space	at the
identity of ${\mathcal T}$ with respect to the Weil-Petersson
K\"{a}hler metric is given by	the following formula for $k\geq 2$: 
\[
{\mathcal A}(e_{k})={\mathcal A}\left( \frac{z^{k}}{\sqrt{k(k^{2}-1)}}\frac{%
\partial }{\partial z}\right) =\left( \sqrt{2 k(k^{2}-1)}
z^{k-2}\right) dz\mbox {.}
\]
\end{thr}

\begin{proof}
We know that the tangent space of the Universal Teichm\"{u}ller
space ${\mathcal T}$ can be identified with the Beltrami
differential on the 
unit disk, i.e.	with tensors of	the type $\mu $	$\overline{dz}\otimes \frac{
\partial }{\partial z}.$ By using the Poincare metric ${\mathcal P=}\frac{1}{
2\pi }{\mathcal (}1-|z|^{2})^{-2}$ $dz\wedge \overline{dz}$	on the
unit disk we 
can identify canonically the space of	Beltrami differentials $\mu $ $%
\overline{dz}\otimes \frac{\partial }{\partial z}$ with	the space of
quadratic differentials	by the standard	map:
\[
\mu \overline{dz}\otimes \frac{\partial	}{\partial z}\rightarrow 
\mu (1-|z|^{2})^{-2}\left( \overline{dz}\right) ^{\otimes 2}.
\]

So the cotangent bundle	of the Universlal Teichm\"{u}ller space	can be
canonically identified with the	quadratic differentials	on the unit disc
restricted to the unit circle. The above expression should be invariant
under the action of the	group ${\mathrm{PSU}}_{1,1}$ on the	unit
disk. So we 
have that if $\omega =f(z)(dz)^{\otimes	2},$ is	a quadratic diffenerial	then
\[
f(\gamma z)=\left( \frac{d\gamma }{dz}\right) ^{2}f(z),
\]
where $\gamma \in {\mathrm	PSU}_{1,1}.$ 
The inner	product	
$\left\langle \omega_{1},\omega _{2}\right\rangle _{W.P.}$
on the quadratic differentials 
$\omega_{1}=f_{1}(z)(dz)^{\otimes 2}$ and 
$\omega _{2}=f_{2}(z)(dz)^{\otimes 2}$
defined	by the Weil-Petersson metric is	given as follows:
\begin{eqnarray*}
\left\langle \omega _{1},\omega _{2}\right\rangle _{W.P.}
 & = & \frac{1}{2\pi i}  
{\int_{\Di_{0}} }(1-|z|^{2})^{4}f_{1}(z)\overline{f_{2}(z)}%
(1-|z|^{2})^{-2}dz\wedge \overline{dz}  \\
& = & \frac{1}{2\pi i}
{\int_{\Di_{0}} }(1-|z|^{2})^{2}f_{1}(z)\overline{f_{2}(z)}dz\wedge
\overline{dz} \mbox{.} 
\end{eqnarray*}

An easy	computation shows that
\begin{equation}
\left\| z^{n}\right\| _{W.P.}^{2}=\frac{1}{2\pi i }
\int_{\Di_{0} }(1-|z|^{2})^{2}|z|^{2n}dz\wedge \overline{dz}=
\int_{0}^{1}(1-r^2)^{2}r^{2n+1}dr,
\end{equation}
where $z=r\exp( i \theta).$ So by integrating twice by
parts we get that
$$\int_{0}^{1}(1-r^2)^{2}r^{2 n + 1}dr=\frac{1}{2(n+1)(n+2)(n+3)}.$$

From	here, we deduce that under the canonical identification between the
tangent	and cotangent bundle on	the universal Teichm\"{u}ller space 
$ \mathrm{Diff}_{+}(S^{1})/{\mathrm PSU}_{1,1}$ is given by the
following formulas 
\begin{multline*}
\Gamma (\mathrm{Diff}_{+}(S^{1})/{\mathrm
PSU}_{1,1},T_{\mathrm{Diff}_{+}(S^{1})/{\mathrm	PSU}_{1,1}}) \\
\longrightarrow  \Gamma \big(\mathrm{Diff}_{+}(S^{1})/{\mathrm{PSU}}_{1,1},
T_{\mathrm{Diff}_{+}(S^{1})/{\mathrm {PSU}}_{1,1}}^{\ast}\big) 
\end{multline*}
$$
\frac{z^{n}}{\sqrt{n(n^{2}-1)}}  \longmapsto  
\sqrt{2 n(n^{2}-1)} z^{n-2}. 
$$
Our Theorem\ \ref{t19} is proved.
\end{proof}

\section{Conclusion of the Proof of the Embedding of ${\mathcal T}$
into the Grassmannian}

As announced in Section~\ref{sec2} we shall now conclude the proof
that ${\mathcal T}$ is embedded in the Segal-Wilson Grassmannian.
This will be done by first showing that if $\|t\|_{2}$ is sufficiently
small, then Assumptions 1 and 2 of Sections~\ref{sec2} and \ref{sec3}
are satisfied, and then by using that the manifold ${\mathcal T}$ is
constructed by right translation of neighborhoods of the identity by
$C^{\infty}$ diffeomorphisms of the circle.  Before that, we start
with some notation and general remarks.

For ${\mathcal W}$ a subspace of $L^{2}(S^{1})$ define
\[ {\mathcal W} \circ \phi \bydef \{ h\circ \phi | 
h \in {\mathcal W} \} \mbox{.} \]

Note that if $\phi$ is a diffeomorphism of class $C^{1}$, then the map
\[ R_{\phi} :  L^{2}(S^{1}) \ni f \mapsto f\circ \phi \in L^{2}(S^{1}) \mbox{ }
\] is a bounded linear transformation. We shall show that if 
$\phi \in C^{2+\epsilon}$ then $R_{\phi}$ maps the Segal-Wilson
Grassmannian into itself. More precisely, $R_{\phi}$ belongs to the
group ${\rm GL}_{\rm res}(L_{2}(S^{1}))$.  We recall \cite{psegal} the
definition of ${\rm GL}_{\rm res}(L_{2}(S^{1}))$. A linear operator $A
\in {\rm GL}_{\rm res}(L_{2}(S^{1}))$ iff the following conditions are
satisfied:

\begin{enumerate}
\item 
$A$ is an invertible bounded linear operator of $L^{2}(S^{1})$ onto 
itself.
\item
If we write $A$ in block matrix form with respect to
the decomposition $L^{2}(S^{1}) = H_{+}\oplus H_{-}$ as
\begin{equation}
 A = \left[
\begin{array}{cc} a &  b \\
                  c &  d \end{array} \right] \mbox{,} 
\label{block}
\end{equation}
then, the operators $b$ and $c$ are Hilbert-Schmidt operators.
\end{enumerate}

As in Section~\ref{sec2}, we use the notation
\[ \mu_{\phi}(z) =
\begin{cases}
\muphi, & \quad  z \in \Di_{0} \\
0, & \quad  z \in {\nEW C} \setminus \Di_{0},
\end{cases} 
\]
where $(f_{0},f_{\infty})$ is a  solution of the normalized
sewing problem associated to $\phi$. (Note that $\phi \in \Dtp$ and hence
we can always choose the representative which is normalized.)

\begin{lm}
\label{nbhd}
Let $\sigma_{t}$ be as defined in Step~2 of the construction of 
Section~\ref{sec4}. Then, $\sigma_{t}$ is quasisymmetric and
there exists $\epsilon>0$ such that 
$ \|t\|_{2} < \epsilon  $ implies 
\begin{equation}
\label{eq1}
\| \mu_{\sigma_{t}} \|_{L^{\infty}} < 1 \mbox{.}
\end{equation}
Let $(f_{0,t},f_{\infty,t})$ denote the solution of the sewing problem
associated to $\sigma_{t}$.
Then, there exists $\epsilon' > 0 $ such that
$ \|t\|_{2} < \epsilon'$ implies that there exists $M$ such that
\begin{equation} 
\label{eq2} 
| {\mathcal S}[f_{\infty,t}(1/z)] | < M  \mbox{,  } \forall z  
\in \Di_{\infty} \mbox{.} 
\end{equation}
Hence, for all $t$ in a sufficiently small neighborhood of
$0$ the Assumptions 1 and 2 are satisfied.
\end{lm}

\begin{proof}
The fact that $\sigma_{t}$ is quasisymmetric is a simple consequence of 
the fact that it is a  $C^{2+\epsilon}$ diffeomorphism, as we remarked 
above.
To prove inequalities (\ref{eq1}) and (\ref{eq2}), we shall
apply the Ahlfors-Weill result, which we review in the Appendix~2.
Notice that if we define 
\[ \varphi(z) \bydef \left( \sum_{k \ge 2} \frac{t_{k} 
(1/z)^{k-2}}{\sqrt{k(k^{2}-1)}} \right)
\left(-\,\frac{2}{z^{4}}\right) \mbox{,} \]
then $\varphi$ is holomorphic inside $\Di_{\infty} $. 
Furthermore, for 
\[ \|t\|_{2} < \epsilon_{0} \mbox{,} \]
where $\epsilon_{0}$ is sufficiently small,  
we have that
\[ \sup_{z \in \Di_{\infty} }
\bigg| -2 \frac{(1-|z|^{2})^{2}}{z^{4}} \sum_{k\ge 2} 
\frac{t_{k} (1/z)^{k-2}}{\sqrt{k(k^{2}-1)}} \bigg|  < 2 \mbox{.} \]

Let's consider the Beltrami equation
\begin{equation}
\label{above}
 \dbar F - \mu_{t} \partial F = 0 \mbox{ ,} 
\end{equation}
where $\mu_{t}$ is the Beltrami differential 
\[ \mu_{t} = 
\begin{cases}
-\frac{1}{2} \frac{(1-|z|^{2})^{2}}{\bar{z}^{4}} \varphi(1/\bar{z}) 
= (1-|z|^{2})^{2} \sum_{k \ge 2} \frac{t_{k} 
\bar{z}^{k-2}}{\sqrt{k(k^{2}-1)}},  & \quad  z\in \Di_{0} \\
 0,  & \quad   z \in \Di_{\infty}. 
\end{cases}
 \]
Because of the Ahlfors-Weill result we have 
\[ \varphi(z) = {\mathcal S}[F] \mbox{ , } z \in \Di_{\infty} \]
where $F$ is a quasiconformal mapping satisfying the 
Beltrami equation~(\ref{above}).
Now, by the construction of $\sigma_{t}$, we have that 
the pair $(f_{0,t},f_{\infty,t})$ is the unique 
solution of the sewing problem for $\sigma_{t}$. Furthermore,
$f_{\infty,t}$ coincides with $F$ on $\Di_\infty$ provided
$F$ is given the normalization (\ref{eqnorm}).

Therefore, 
\[ \varphi(z)  = {\mathcal S}[f_{\infty,t}](z) \mbox{, } 
z \in \Di_{\infty} \mbox{.} \]
In other words,
\[ {\mathcal S}[f_{\infty,t}](z) 
= \left(-\frac{2}{z^{4}}\right) \sum_{k \ge 2}
\frac{t_{k} (1/z)^{k-2}}{\sqrt{k(k^{2}-1)}} 
\mbox{, }  z \in \Di_{\infty} \mbox{.} \]
Taking $\|t\|_{2}$ sufficiently small we then have 
\[ \left\|-\frac{1}{2} (1-|z|^{2})^{2} {\mathcal S}
[f_{\infty,t}(1/z)] \right\|_{L^{\infty}(\Di_{0})}
\le c_{0} < 1 \mbox{,} \] 
and that
${\mathcal S}[f_{\infty,t}(1/z)] $ is bounded in $\Di_{0}$.
\end{proof}

The goal now is to extend the embedding of the neighborhood of the
identity of ${\mathcal T}$ in the Grassmannian to all the manifold
${\mathcal T}$.  To perform this extension we will need the following:

\begin{lm}
\label{explm}
Let $\phi \in \Dinfty$ and $\sigma_{t}$ in the image of the exponential map
for $t$ in a sufficiently small neighborhood of $0$ so that
Assumptions 1 and 2 are satisfied.
Take 
\[\chi \bydef \mu_{\sigma_{t}\circ \phi} \]
 and 
\[ \nu \bydef \mu_{\sigma_{t}} \mbox{.} \] 
Then, 
\[ \ker \dbar_{\chi} \bigg|_{S^{1}}  = 
 \ker \dbar_{\nu} \bigg|_{S^{1}} \circ \phi \mbox{.} \]
\end{lm}

\begin{proof}
Take $\psi=\sigma_{t} \circ \phi$.
Let $\widetilde{\psi}$ and $\widetilde{\sigma_{t}}$ be the Beurling-Ahlfors
extensions of $\psi$ and $\sigma_{t}$, respectively. 
We are going to show that $\widetilde{\psi}^{-1}\circ \widetilde{\sigma_{t}}$
maps solutions of 
\begin{equation}
 \dbar_{\chi} G = 0 
\label{eqpsi}
\end{equation}
into solutions of
\begin{equation}
 \dbar_{\nu} \widetilde{G} = 0
\label{eqsig}
\end{equation}
by means of 
\[ \widetilde{G} = G \circ (\widetilde{\psi}^{-1} \circ \widetilde{\sigma_{t}})
\mbox{.} \]
Let $F_{\chi}$ and $F_{\nu}$ be the unique solutions of 
equations (\ref{eqpsi}) and (\ref{eqsig}) with the conditions of
say fixing the points $-1$, $-i$, and $0$. (Theorem~\ref{exist} of Appendix~2.)

\medskip\noindent
{\it Claim.} \  
For points in the interior of $\Di_{0}$ we have
\[ \dbar_{\nu} F_{\chi}\circ \widetilde{\psi}^{-1} 
\circ \widetilde{\sigma_{t}} = 0 \mbox{.} \]
Indeed, since $F_{\chi}$ and $\widetilde{\psi}$ both have the same Beltrami
coefficient inside $\Di_{0}$ it follows that 
$\dbar F_{\chi} \circ \widetilde{\psi}^{-1} = 0$.
So, $F_{\chi}\circ \widetilde{\psi}^{-1}$ is analytic inside $\Di_{0}$.
The claim follows since 
$ \dbar_{\nu} f\circ \widetilde{\sigma_{t}} = 0$ inside $\Di_{0}$ for
any analytic function $f$ and because $\widetilde{\sigma_{t}}$
maps $\Di_{0}$ on itself. 

Take now $G$ any solution of (\ref{eqpsi}), then by Claim~($\beta$)
of page 258 of \cite{bersjohn} it follows that
there exists an analytic function $g$ such that
$G = g \circ F_{\chi}$. The same argument employed to prove the
Claim gives that  inside $\Di_{0}$ 
\[ \dbar_{\nu} g\circ f \circ \widetilde{\sigma_{t}} = 0 
\mbox{.} \]
For points $z\in S^{1}$ we have that 
$\widetilde{\psi}^{-1}\circ \widetilde{\sigma_{t}} (z) = \phi^{-1}(z) $.
We conclude the proof by remarking that the function
$\widetilde{\psi}^{-1}\circ \widetilde{\sigma_{t}}$ admits a quasiconformal 
extension $\Gamma$ to $\widehat{\nEW C}$ with 
$\dbar \Gamma = 0 $ in $\Di_{\infty}$.
Hence, if $\dbar_{\chi}G =0$ in $\Di_{\infty}$,
then $\widetilde{G} = G \circ \Gamma$ is a solution of 
$\dbar_{\nu} \widetilde{G} = 0$. 
The continuity of $\Gamma$ on the boundary and the fact that
for $z\in S^{1}$ we have
\[ \phi^{-1}(z) = \widetilde{\psi}^{-1}\circ \widetilde{\sigma_{t}} (z) =
\Gamma(z) \]
implies that
\[   \ker \dbar_{\nu} \bigg|_{S^{1}} \supset 
\ker \dbar_{\chi} \bigg|_{S^{1}} \circ \phi^{-1} \mbox{.} \] 
A similar argument writing $\sigma_{t}=\sigma \circ \phi^{-1}$, and using
$\phi$  in the role of $\phi^{-1}$ gives the 
equality. 
\end{proof}

As a consequence of the previous Lemma it follows that every point in the 
manifold ${\mathcal T}$ is associated to an element of the Grassmannian $Gr$.
More precisely, we take $\sigma \in {\mathcal T}$ and write
$\sigma=\sigma_{t} \circ \phi$ with $t$
in a sufficiently small neighborhood provided by Lemma~\ref{nbhd}.
For $\sigma_{t}$ we know from Theorem~\ref{th8} 
that the space ${\mathcal W}_{\sigma_{t}}$ of Definition~\ref{crucial} is a 
point in the Grassmannian.  We now define
\[ {\mathcal W}_{\sigma} = {\mathcal W}_{\sigma_{t}} \circ \phi \mbox{.} \]

\begin{thr}			
Let $\sigma \in {\mathcal T}$, then ${\mathcal W}_{\sigma}$  
is a point of the Segal-Wilson Grassmannian ${\rm Gr}$.
\end{thr}

\begin{proof} 
We write $\sigma=\sigma_{t} \circ \phi$ with $t$
in a sufficiently small neighborhood provided by Lemma~\ref{nbhd}.
Note that $\ker \dbar_{\mu_{\sigma_{t}}}\Big|_{S^{1}}$ is dense
in ${\mathcal W}_{\sigma_{t}}$.
From the Lemma~\ref{explm} we have
that 
\[ \ker \dbar_{\mu_{\sigma}}\Big|_{S^{1}} = 
R_{\phi}\left( \ker \dbar_{\mu_{\sigma_{t}}}\Big|_{S^{1}}\right) \mbox{,} \]
where $R_{\phi}$ was defined in the beginning of the present section.
The proof reduces to showing that $R_{\phi}$ belongs to the
group ${\rm GL}_{\rm res}(L_{2}(S^{1}))$,
Now, the operator 
\[ R_{\phi}:L^{2}(S^{1})\rightarrow L^{2}(S^{1}) \]
can be written as
\[ R_{\phi} = M_{\phi} \cdot C_{\phi} \mbox{,} \]
where
\[ M_{\phi}[ f ](z) \bydef \frac{f(z)}{\sqrt{\phi'(z)}}  \mbox{,} \]
and 
\[ C_{\phi}[f](z) \bydef   f\circ\phi^{-1} (z) \sqrt{\phi'(z)}
\mbox{.} \]
Since $\phi\in \Dinfty$, it is shown on page 91 of \cite{psegal} (see
also \cite{segal2}) that $C_{\phi}$ belongs to 
${\rm GL}_{\rm res}(L_{2}(S^{1}))$.
As for $M_{\phi}$ 
the difficulty lies on the fact that the loop
\[ z \mapsto  (1/\sqrt{\phi'(z)}) \] 
is not necessarily
continuous. However, it has finitely many discontinuity points.
An easy adaptation of the argument on page 83 of \cite{psegal} gives
that $M_{\phi}$ also belongs to ${\rm GL}_{\rm res}(L_{2}(S^{1}))$.
An alternative way is to follow directly the method of \cite{psegal}, namely, 
to consider the operator  
\[ J :  H_{+}\oplus H_{-} \ni (f_{+},f_{-})  \mapsto (f_{+},-f_{-})
\in H_{+}\oplus H_{-} \mbox{,}  \]
and to show that the kernel of the integral operator 
that represents the commutator $[R_{\phi},J]$ is Hilbert-Schmidt.
This shows that the off-diagonal terms of the decomposition of $R_{\phi}$
in the form (\ref{block}) are Hilbert-Schmidt operators.
\end{proof}

\medskip\noindent
{\bf Remark.} \ The map
$\Pi:{\mathcal T}^{\infty} \rightarrow Gr$, where
\[ \Pi: \sigma \mapsto N_{\sigma}=\mbox{\rm span}\{1,\sigma,\sigma^{2},\dots \}
\mbox{.}\]
was defined in \cite{nag}.
Based on the argument of this section it is easy to see that this map 
coincides with
our map $\sigma \mapsto W_{\sigma}$ restricted to ${\mathcal T}^{\infty}$.
The proof is a consequence of the
Lemma~\ref{explm} of Section~\ref{sec4p5}.

\label{sec4p5}
\section{The Curvature Computation}

In this section we show that the Teichm\"uller space ${\mathcal T}=$
\linebreak
$\Dtp $ equipped with the unique invariant K\"{a}hler metric has
negative curvature in holomorphic directions.\footnote{ After a first
draft of this article was written we learned of the work of Misiolek
\cite{misiolek}, where formulae for the curvature are also given. We
remark, however, that he is working in a different orbit of the
Bott-Virasoro group.}  More precisely, we will show that the curvature
is negative and uniformly bounded away from zero in holomorphic
directions. This combined with an extension of the Hopf-Rinow Theorem
will be used in the next section to yield the existence of geodesics
for all time.

In the Appendix~1 we will prove that the geodesics of 
${\mathcal T} $ yield solutions to the Korteweg-de Vries
equation. See also G. Segal's paper \cite{segal}.

We start reviewing a few facts about the 
invariant K\"ahler metric of the Grassmannian.
We follow closely the exposition and notation of Section~7.8 of \cite{psegal}.
The first step in the construction of the K\"ahler metric is to
define it at the point $H_{+} \equiv H_{+}\oplus\{0 \} \in \mbox{\rm Gr}$.
Note that 
\[T_{H_{+}} \mbox{\rm Gr} = HS(H_{+},H_{-}) \mbox{ ,} \]
where $ HS(H_{+},H_{-})$ denotes the space of Hilbert-Schmidt
operators from $H_{+}$ into $H_{-}$.
Hence, it is natural to define\footnote{Our definition differs from
the one used in \cite{psegal} by a factor of 2.}
for $\psi$ and $\chi$ in $T_{H_{+}} \mbox{\rm Gr}$
\[ \langle \psi , \chi \rangle \bydef  \mbox{\rm Tr}(\psi^{\ast} \chi)
\mbox{.} \]
To extend the definition to the rest of the Grassmannian one
uses the fact that $GL_{res}(H)$ acts transitively on $\mbox{\rm Gr}$.
(in fact $U_{res}(H)$ already acts transitively). Let $A \in GL_{res}(H)$ be a 
transformation sending $H=H_{+}\oplus H_{-}$ into $H=W\oplus W^{\perp}$,
and preserving the direct sum decomposition.
Then, if $\widetilde{\psi}$ and  $\widetilde{\chi}$ are elements of
$T_{W} \mbox{\rm Gr} \equiv HS(W,W^{\perp})$ we have that
$\psi \bydef A^{-1} \widetilde{\psi} A$ and 
$\chi \bydef A^{-1} \widetilde{\chi} A$ belong to $T_{H_{+}} \mbox{\rm Gr}$.
Hence, to have a right-invariant metric we must set
\begin{eqnarray} 
\langle \widetilde{\psi},\widetilde{\chi} \rangle_{W} &  = & 
\langle \psi , \chi \rangle_{H_{+}} \\
& = & \mbox{\rm Tr} (A^{\ast} \pst^{\ast}(A A^{\ast})^{-1} \cht A ) \\
& = & \mbox{\rm Tr} (\pst^{\ast}(A A^{\ast} )^{-1} \cht A A^{\ast}) \mbox{,} 
\label{vareq}
\end{eqnarray} 
where in the last equality we used a well known fact. Namely, that 
$\mbox{\rm Tr}(AB)= \mbox{\rm Tr}(BA)$ for any $A$ bounded and 
$B$ trace-class. Note that as a  particular case, 
if $A\in U_{res}(H)$ then 
$\langle \widetilde{\psi},\widetilde{\chi} \rangle_{W} 
= {\rm Tr} (\pst^{\ast}\cht)$.

It was shown in \cite{nag} that the embedding of ${\mathcal
T}^{\infty}$ into the Grassmannian is isometric.  Since $T_{id}
{\mathcal T}^{\infty}$ is dense in $T_{id} {\mathcal T}$, and our
embedding coincides with that of \cite{nag} in ${\mathcal
T}^{\infty}$, it follows that the embedding of ${\mathcal T}$ into the
Grassmannian we obtained is also isometric.  Here, however, we have at
both sides Hilbert manifolds.

\begin{thr}				
Let ${\mathcal T}=\Dtp$ be 
equipped with
the unique right-invariant K\"{a}hler metric and isometrically embedded
in the Grassmannian $\mbox{\rm Gr}$.
Let $\{\psi_{k}\}$ be an orthonormal basis of 
\[T_{H_{+}} \mbox{\rm Gr} = HS(H_{+},H_{-}) \mbox{.} \]
Then,
\begin{itemize}
\item[{\rm (a)}] The component of the curvature tensor with
respect to the orthonormal basis  $\{\psi_{k}\}$ 
is given by:
\begin{multline*} 
R_{i\bar{j},k\bar{l}} = - \delta_{i\bar{j}} \delta_{k\bar{l}}
- \delta_{i\bar{l}} \delta_{k\bar{j}}  \\
+ {\rm Tr}(\psi_{i}\psi_{j}^{\ast} \wedge \psi_{k}\psi_{l}^{\ast}) + 
{\rm Tr}(\psi_{i}\psi_{l}^{\ast} \wedge \psi_{k}\psi_{j}^{\ast})\; 
\mbox{ for } (i,j)\ne (k,l) 
\end{multline*}
and
\[ R_{i\bar{j},i\bar{j}} = - 2  \delta_{i\bar{j}} +
{\rm Tr}(\psi_{i}\psi_{j}^{\ast} \wedge \psi_{i}\psi_{j}^{\ast} )\mbox{.}\]

\item[{\rm (b)}]
For  any complex direction $\psi$, if we denote by $K_{\psi}$ the
Gaussian sectional curvature in the two-dimensional
space defined by ${\rm Re} \psi$ and ${\rm Im} \psi$,
we have 
\[ K_{\psi} =-2 + \mbox{\rm Tr}(\psi\psi^{\ast} 
\wedge \psi\psi^{\ast}) <  -3/2 \mbox{.} \]

\end{itemize}
\end{thr}

\proof
The proof is based on the construction of the so-called
Cartan coordinate system \cite{GrifHarris}.
By that we mean a holomorphic  coordinate system 
$(x^{1},x^{2},\dots)$ in which the components of the
metric tensor $g_{i,\bar{j}}$ is given by the
formula:
\[g_{i,\bar{j}}=\delta_{i,\bar{j}} +
r_{i \bar{j},k \bar{l}} x^{k} \bar{x}^l + {\mathcal O}(|x|^{3}) 
\mbox{.} \]
Cartan \cite{royden} proved 
that if the coordinate system satisfies this last  equation,
then 
\[ r_{i \bar{j},k \bar{l}}= -R_{i \bar{j},k \bar{l}}  \mbox{,}\]
where $R$ is the curvature tensor.

\medskip\noindent
{\it Construction of the Cartan's coordinates.}

Define the exponential map as follows:
\footnote{ This is the same map constructed by Nag in \cite{nag}.}
\[ \exp: HS(H_{+},H_{-}) \rightarrow \mbox{\rm Gr} \mbox{.}  \]
We assign to $\varphi_{t}= \sum t_{i}\psi_{i} \in HS(H_{+},H_{-})$,
the subspace $W_{t}$ of $H$ spanned by the set
\[ \{ 1+\varphi_{t}(1),z+\varphi_{t}(z),\dots,z^{n}+\varphi_{t}(z^{n}),
\dots \} \]
Obviously, $W_{t}$ is the graph of the operator $\varphi_{t}$. 
Here, $t=(t_{1},t_{2},\dots)$ defines the local coordinates.

\begin{lm} \label{lem7}			
In the above coordinates the following expansion near \linebreak
$t=0$  holds 
\begin{align*}
& -\frac{\partial^{2}}{\partial t_{i} \partial \bar{t}_{j} }
\log \det (id - \sum t_{i} \bar{t}_{j} \psi_{i}\psi^{\ast}_{j})\\  
& = \delta_{i\bar{j}}+(2 \delta_{i\bar{j}} - 
{\rm Tr}(\psi_{i}\psi_{j}^{\ast} \wedge \psi_{i}\psi_{j}^{\ast})) 
t_{i}\bar{t}_{j}\\
&\quad +\!\! \sum_{(k,l)\ne(i,j)}\!\left[\delta_{i\bar{j}}\delta_{k\bar{l}}
+ \delta_{i\bar{l}} \delta_{k\bar{j}}  -
{\rm Tr}(\psi_{i}\psi_{j}^{\ast} \wedge \psi_{k}\psi_{l}^{\ast})  - 
{\rm Tr}(\psi_{i}\psi_{l}^{\ast} \wedge \psi_{k}\psi_{j}^{\ast}) \right]
t_{k}\bar{t}_{l} \\
&\quad +  {\rm h.o.t.}
\end{align*}
\end{lm}

\begin{proof}
Let
\begin{equation}
 f(t)=\det \left(id-\sum t_{i}\bar{t}_{j}\psi_{i}\psi^{\ast}_{j}\right)
\mbox{,} \label{eq9}
\end{equation}
then
\begin{equation}
 \frac{\partial^{2}}{\partial t_{i} \partial \bar{t}_{j} }
\log f = 
\frac{ \partial^{2} f}{\partial t_{i} \partial \bar{t}_{j} } f^{-1}
- \frac{ \partial f}{\partial t_{i}}\frac{ \partial f}{\partial \bar{t}_{j}}
f^{-2}  \mbox{.}
\label{eq:eq1}
\end{equation}
From the definition of the determinant, we  have that 
\begin{align*} 
& \det \left(id - \sum t_{i} \bar{t}_{j} \psi_{i}\psi^{\ast}_{j}\right)\\
& =1-\sum_{i,j}  t_{i} \bar{t}_{j} {\rm Tr}(\psi_{i}\psi^{\ast}_{j})
+ \sum_{i,j,k,l}  t_{i} \bar{t}_{j}  t_{k} \bar{t}_{l}
{\rm Tr}(\psi_{i}\psi^{\ast}_{j} \wedge  \psi_{k}\psi^{\ast}_{l}) 
+ {\rm h.o.t.} 
\end{align*}
Hence,
\begin{align*}
& \frac{\partial^{2}}{\partial t_{i} \partial \bar{t}_{j} }
\det \left(id - \sum t_{i} \bar{t}_{j}
\psi_{i}\psi^{\ast}_{j}\right)\\
& = -\delta_{i\bar{j}} + 
{\rm Tr}(\psi_{i}\psi_{j}^{\ast} \wedge \psi_{i}\psi_{j}^{\ast}))
t_{i}\bar{t}_{j} \\
&\quad + \sum_{(k,l)\ne(i,j) } {\rm Tr}\left[ 
\psi_{i}\psi^{\ast}_{j} \wedge  \psi_{k}\psi^{\ast}_{l} +  
 \psi_{i}\psi_{l}^{\ast} \wedge \psi_{k}\psi_{j}^{\ast}\right]
t_{k} \bar{t}_{l} + {\rm h.o.t.}  \mbox{,}
\end{align*}
\begin{align*}
\frac{\partial}{\partial t_{i} }\det \left(id - \sum t_{i} 
\bar{t}_{j} \psi_{i}\psi^{\ast}_{j}\right) & =
 - \sum_{l} {\rm Tr}(\psi_{i}\psi^{\ast}_{l}) \bar{t}_{l} + {\rm h.o.t.}
\mbox{,} \\
\intertext{and} 
\frac{\partial}{\partial \bar{t}_{j} }\det \left(id - \sum t_{i} 
\bar{t}_{j} \psi_{i}\psi^{\ast}_{j}\right) & =
 - \sum_{k} {\rm Tr}(\psi_{k}\psi^{\ast}_{j}) t_{k} + {\rm h.o.t.}
\end{align*}
Furthermore,
\[ 1/f=1+\sum t_{i} \bar{t}_{j}{\rm Tr}( \psi_{i}
\psi^{\ast}_{j})+ {\rm  h.o.t.} \]
and
\[ 1/f^{2}=1+ 2 \sum t_{i} \bar{t}_{j}{\rm Tr}
( \psi_{i}\psi^{\ast}_{j}) + {\rm  h.o.t.}\]

Substituting the last 5 equations into equation  (\ref{eq:eq1})
we obtain the result. 
This completes the proof of the lemma.
\end{proof}

From \cite{nag} it follows that $\log f(t)$, with $f$ defined by  (\ref{eq9}),
is the potential of the K\"{a}hler metric.
Hence, Lemma~\ref{lem7} gives that the coordinates 
$(t_{1},t_{2},\dots)$ forms 
a Cartan coordinate system. Thus,   for $(i,j)\ne (k,l)$ we get:
\[ R_{i\bar{j},k\bar{l}} = - \delta_{i\bar{j}} \delta_{k\bar{l}}
- \delta_{i\bar{l}} \delta_{k\bar{j}}  +
{\rm Tr}(\psi_{i}\psi_{j}^{\ast} \wedge \psi_{k}\psi_{l}^{\ast} + 
  \psi_{i}\psi_{l}^{\ast} \wedge \psi_{k}\psi_{j}^{\ast} ) \mbox{,}\]
and 
\[ R_{i\bar{j},i\bar{j}} = - 2  \delta_{i\bar{j}} +
{\rm Tr}(\psi_{i}\psi_{j}^{\ast} \wedge \psi_{i}\psi_{j}^{\ast} )\mbox{.}\]

This concludes the proof of part~(a) of the theorem.
To prove part~(b) remark that the Gaussian curvature in the
direction $\psi_{i}$ is given by
\[ K_{\psi_{i}}=R_{i\bar{i},i\bar{i}}= -2 + 
{\rm Tr}(\psi_{i}\psi_{i}^{\ast} \wedge \psi_{i}\psi_{i}^{\ast}) \mbox{.}\]
Now, if $\psi_{1}=\varphi $, we complete the set $\{\psi_{1}\}$ to 
an orthonormal set in the Hilbert space $T_{H_{+}}\mbox{\rm Gr}$.
Hence, from the previous discussion it follows that
\[ K_{\varphi} = -2 + {\rm Tr}(\varphi \varphi^{\ast}
\wedge \varphi \varphi^{\ast}) \mbox{.}\] 

\begin{lm}				
If ${\rm Tr}(\varphi \varphi^{\ast})=1$, then 
${\rm Tr}( \wedge^{2} \varphi \varphi^{\ast}) <  1/2.$
\end{lm}

\begin{proof}
Note that $\varphi \varphi^{\ast}$ is a compact positive 
operator and hence its nonzero eigenvalues are all positive.
Let $\{\lambda_{i}\}$ be the set of such nonzero
eigenvalues. Since $\varphi \varphi^{\ast}$ is trace-class
and $\|\varphi\|^{2}={\rm Tr} \varphi \varphi^{\ast} = 1$
it follows that
${\rm Tr} \varphi \varphi^{\ast} = \sum \lambda_{i} = 1  $.
A simple argument with tensor products gives
\[ Tr(\wedge^{2} \varphi \varphi^{\ast}) = \sum_{i <  j} 
\lambda_{i} \lambda_{j} \mbox{.} \]
Now we remark that
\[ 1 = \left( \sum \lambda_{i} \right)^{2} 
= 2 \sum_{i < j} \lambda_{i} \lambda_{j}
+ \sum \lambda_{i}^{2} \mbox{.}\]
From here it follows that
\[ Tr(\wedge^{2} \varphi \varphi^{\ast}) < 1/2 \mbox{.} \]
Using the formula for the curvature in holomorphic direction $\varphi$
it follows that
\[ K_{\varphi} < -3/2 < 0 \mbox{.} \]
\end{proof}

\label{sec5}
\section{Description of the Geodesics}			

In this section we shall give a description 
of the geodesics passing through the identity of 
${\mathcal T} \hookrightarrow  Gr$
in the complex direction  $ \psi $.
We recall from Section~\ref{sec5}
the definition of the exponential map.
It was defined, for each $s\in {\nEW C}$ and $\psi \in HS(H_{+},H_{-})$ as the 
subspace of $H$ spanned by the set of vectors
\[\{1+s\psi(1),z+s\psi(z), \dots \} \mbox{.} \]

Our immediate goal is to show that:

\begin{thr}				
The complex curve $\gamma_{\psi}(s)=\exp(s\psi)$ is
a totally geodesic 2-real-dimensional submanifold
of ${\mathcal T}$.
\end{thr}

We recall that a submanifold $S$ of a Riemannian manifold $M$ is called
geodesic at $p$ if each $M$-geodesic passing through $p$ in
a tangent direction to $S$ remains in $S$ for all time. If 
$S$ is geodesic at all its points, 
then it is called totally geodesic~\cite{helgason}.

The proof of the Theorem is a consequence of the following two results:

\begin{lm}				
Let $\psi \in HS(H_{+},H_{-})$ be such that $\|\psi\|^2=1$
and $s(t)=s_{0}+e^{i \theta} t$, with $\theta \in {\nEW R}$.
Then, the norm of the tangent vector $\dot{\gamma}_{\psi}$
to the the path $t\mapsto \gamma_{\psi}(s(t))$ is given by 
\[ \| \dot{\gamma}_{\psi} \| = 1 \mbox{.} \]
\end{lm}

\begin{proof}  
Call ${\mathcal W}_{s}$ the subspace $\gamma_{\psi}(s) \subset H$. 
With respect to the decomposition $H=H_{+}\oplus H_{-}$ 
define the block matrix  
\[ A_{s} = \left[ 
\begin{array}{cc} id & - \bar{s} \psi^{\ast} \\
		  s \psi  &   id \end{array} \right]. \]
From the definition of ${\mathcal W}_{s}$ we have
\[ {\mathcal W}_{s} = A_{s} H_{+} \mbox{.} \]
Let $T:H\rightarrow H$ be any bounded operator. Using that 
$\mbox{\rm Graph} T$ is perpendicular to 
$\mbox{\rm Graph}' T^{\ast}$, where
\[ \mbox{\rm Graph}'T \bydef \{(-Tx,x) \ | \  x \in 
\mbox{\rm dom}(T) \}  \mbox{,}  \]
we get  
\[ {\mathcal W}_{s}^{\perp} = A_{s} H_{-} \mbox{.} \]
Hence, the matrix $A_{s}$ maps $H=H_{+}\oplus H_{-}$ into 
$W_{s}\oplus W_{s}^{\perp}$ and preserves the direct sum
decomposition.

It is well known that the operators $ (id +  s \bar{s} \psi \psi^{\ast})$
and $(id +  s \bar{s} \psi^{\ast} \psi)$ are invertible. The
following relations can be checked easily:
\begin{equation}
\label{eqa1} 
(id +  s \bar{s} \psi^{\ast} \psi)^{-1}
\psi^{\ast} =  \psi^{\ast} (id +  s \bar{s} \psi \psi^{\ast})^{-1} \mbox{,} 
\end{equation}
and
\begin{equation}
\label{eqa2} 
(id +  s \bar{s} \psi \psi^{\ast})^{-1}
\psi =  \psi(id +  s \bar{s} \psi^{\ast} \psi)^{-1} \mbox{.} 
\end{equation}
Therefore, 
\begin{align} 
A_{s}^{-1} & =  \left[
\begin{array}{cc} id &  \bar{s} \psi^{\ast} \\
                  -s \psi  &   id \end{array} \right]
\left[ \begin{array}{cc} (id +  s \bar{s} \psi^{\ast} \psi)^{-1} & 0 \\
0 & (id +  s \bar{s} \psi \psi^{\ast})^{-1}
                   \end{array} \right]  \\[5pt]
 & =  \left[ \begin{array}{cc} (id +  s \bar{s} \psi^{\ast} \psi)^{-1} & 0 \\
0 & (id +  s \bar{s} \psi \psi^{\ast})^{-1}
                   \end{array} \right] 
 \left[
\begin{array}{cc} id &  \bar{s} \psi^{\ast} \\
                  -s \psi  &   id \end{array} \right] \mbox{.} 
\label{eqa3}
\end{align}
To compute the norm $\|\dot{\gamma}_{\psi}(s)\|$, we use the identification 
of $T_{{\mathcal W}_{s}} \mbox{\rm  Gr}$ with 
$HS({\mathcal W}_{s},{\mathcal W}_{s}^{\perp})$. The latter, is mapped onto
$HS(H_{+},H_{-})$ by means of 
\[
\widetilde{X} \mapsto A_{s}^{-1} \widetilde{X} A_{s}  \Big|_{H_{+}} 
\mbox{.} \]
From the invariance of the metric we have that the norm of 
$\widetilde{X}\in T_{{\mathcal W}_{s}} \mbox{\rm  Gr}$
is given by 
\begin{equation}
\label{vareq1} \|\widetilde{X}\|^{2} = \Tr 
 \left( \left(A_{s}^{-1}\widetilde{X} A_{s}\Big|_{H_{+}}\right)^{\ast}
\left(A_{s}^{-1}\widetilde{X} A_{s}\Big|_{H_{+}}\right) \right).
\end{equation} 
Since, 
\[ \dot{A} = 
\left[ \begin{array}{cc} 0 &   -e^{-i\theta} \psi^{\ast} \\
  e^{i\theta} \psi  &   0 \end{array} \right] \mbox{,} \]
a technical but straightforward computation with
$\widetilde{X}=\dot{A}(s)$ taking into account
equations~(\ref{eqa1}), (\ref{eqa2}),
and (\ref{eqa3}) yields  the lemma. 
\end{proof}

\begin{lm}
Set 
\[\dot{\gamma}_{\psi}(s) =  \frac{d}{ds} \gamma_{\psi}(s) \]
and assume that 
\[\|\dot{\gamma}_{\psi}(s)\|^2 = 1 \mbox{.} \]
Let $\nabla$ be the covariant derivative in ${\mathcal T}$
given by the Levi-Civita connection.
Then,
\[ \nabla_{\dot{\gamma}_{\psi}(s)} \dot{\gamma}_{\psi}(s) = 0 \mbox{.}\]
\end{lm}

\begin{proof}
Let $\gig(s)$ be a geodesic in our  K\"ahler manifold.
For each point $s$ of the geodesic $\gig(s)$ we define a complex direction
as follows: 
\[ \dot{\gig}(s)+i {\mathcal I} \dot{\gig}(s)  \mbox{.}\]
For each $s$ in $\gig(s)$ we consider a geodesic ${\mathcal I} \dot{\gig}(s)$ 
and so
each point $\tau$ on the geodesic from $s$ with direction 
${\mathcal I} \dot{\gig}(s)$ we have two tangent 
vectors. The first, $\alpha(\tau)$, which is the parallel transport of 
$\dot{\gig}(s)$
and the second $\beta(\tau)$ which is given by
\[ \beta(\tau)=\frac{d}{ds} {\mathcal I} \dot{\gig}(s) \mbox{,} \]
and is the parallel transport of ${\mathcal I} \dot{\gig}$.
Now, the K\"{a}hler condition gives that 
$[\dot{\gamma}(s), {\mathcal I}\dot{\gamma}(s)]= 0$. Hence, it
follows from Frobenius theorem that there exist a surface $S$
such that the tangent space is spanned by $\alpha(s)$ and 
$\beta(s)$. Since, ${\mathcal I} (\alpha(s) + i \beta(s) )=
i( \alpha(s) + i \beta(s)) $ it follow that $S$ is a complex analytic curve
and we can take $z$ as a complex analytic coordinate associated to
the point $z = \exp (x \dot{\gig}(s)  + i y {\mathcal I} \dot{\gig}(s)) $.
Let's write
\[ \dot{\mu}(z) = \dot{\gig}(s) + i {\mathcal I} \dot{\gig}(s)  \mbox{.} \]

From the properties of the Levi-Civita connection
it follows that
\begin{equation} 
0 = \frac{d}{dz} \| \dot{\mu}(z) \|^{2}\Big|_{s_{0}}
= \langle \nabla_{z} \dot{\mu}(z) ,  \dot{\mu}(z) \rangle = 0  \mbox{.} 
\label{eq:eq0}
\end{equation}
and so
\[ \frac{d^{2}}{d\bar{z}dz} \| \dot{\mu}(z) \|^{2}\Big|_{s_{0}} 
= \|\nabla_{z} \dot{\mu}(z)\|^{2}-R \| \dot{\mu}(z) \|^{2} = 0. \]
Since $R<0$, we have
\[ \nabla_{z} \dot{\mu}(z) = 0  \mbox{.} \]
Restrict $ \nabla_{z} \dot{\mu}(z) $ on the real part at $s_{0}$ 
and we get  that
\[ \nabla_{z} \dot{\gig}(z) = 0 \mbox{.} \]
Hence, taking ${\rm Re} z= \dot{\gig}$, we have  
\[ \nabla_{\dot{\gig}(s)} \dot{\gig}(s) \Big|_{s_{0}} = 0 \mbox{.} \]
\end{proof}

\medskip\noindent
{\bf Remark.} \ 
As a consequence of the previous theorem, we can characterize a
geodesic in the space ${\mathcal T}$ by considering this space 
embedded in the Grassmannian $Gr$. 
Given a real direction $v$  define
\[ \psi = v + i {\mathcal I} v \mbox{,} \]
where ${\mathcal I}$ is the complex structure operator.
Clearly $v$ is  contained in the 2-real-dimensional
plane spanned by $\psi$.
Since $\exp(s \psi)$ is  a totally geodesic submanifold of ${\mathcal T}$
it follows that the geodesic in the real direction $v$ will
be contained in $\{\exp(s \psi) | s \in {\nEW C} \}$.

\medskip
The above remarks allow us to state: 

\begin{thr}\label{alltime}				
For each point $\phi$ in ${\mathcal T}$ and each direction $v$ in the
tangent space $T_{\phi}{\mathcal T}$ the geodesic $\gamma(s)$ passing
through $\phi$ in the direction $v$ exists for all $s \in {\nEW R}$.
\end{thr}

\begin{proof}
Given the homogeneity of the space ${\mathcal T}$ without loss of generality 
we may assume that $\phi$ is the identity map of $S^{1}$.
Because of the completeness of ${\mathcal T}$ it follows that
$\exp(s \psi)$ is a  complete closed submanifold.
Now, we can apply Hopf-Rinow Theorem \cite{docarmo} to the 
{\em finite dimensional}
manifold $\{ \exp(s \psi) | s\in {\nEW C} \}$ to conclude that the
geodesic exists for all time. 
\end{proof} 

\begin{co} \label{andrey}				
The periodic solution to KdV equation with initial data 
\begin{equation}
u_{0}(x) = \sum 
a_{n} e^{inx} \in \HSob^{3/2}({\nEW{ R} },{\nEW{ R} })
\label{condit}
\end{equation}
exists for all time. 
\end{co}

\begin{proof}
Consider the pairing between (real) quadratic forms on $S^{1}$
and (real) vector fields. It is given by
\[ \left\langle q(x) dx^{\otimes 2} | v(x) \frac{d}{dx} \right\rangle 
= \int_{0}^{2\pi} q(x) v(x) dx \]
Using the Weil-Petersson metric we have an identification
between the tangent and cotagent spaces. Let's denote by 
${\mathcal A}:  T_{id} {\mathcal T} \rightarrow T_{id} {\mathcal T}^{\ast}$ 
such identification.
Given an initial condition
$u_{0}$ as in equation~(\ref{condit}) we 
take $\psi = {\mathcal A}^{-1} u_{0}$.  

Let $\gamma_{\psi}(t,z)$  be the geodesic whose existence is
guaranteed for all  real time by Theorem~\ref{alltime}.
From the results in the Appendix~1 it follows that
the function

\[ u(t,x) = {\mathcal A} \left( \frac{\dot{\gig}
(t,\exp(ix))}{\gig'(t,\exp(ix))}\right)  \mbox{,} \]
where $'$ denotes the derivative w.r.t. $z$,
is a solution to the KdV equation. Since
$\gamma_{\psi}(t,\cdot)$ exists for all time, so does $u(t,\cdot)$.
\end{proof}

We close this section with an infinite dimensional analogue 
of Hada\-mard's theorem, which
will be used in the next section to show exponential spreading
of the geodesic flow on ${\mathcal T}$.

\begin{thr}				
The exponential map gives a global diffeomorphism  between the
tangent space $T_{id}{\mathcal T}$ and ${\mathcal T}$.
\label{cartan}
\end{thr}

\begin{proof} 
First we prove that for fixed $\psi$, the exponential map 
restricted to the two-real dimensional space $\{t\psi| t \in {\nEW C}\}$
and taking values in the totally geodesic submanifold $D_{\psi}$ 
is a covering map.
To show this we use that the curvature in the holomorphic direction is
negative, and so it implies that the norm 
\[ \|d \exp(t v) v \| \ge \|v\| \mbox{.} \]
From a standard result in (finite dimensional) differential
geometry it follows that the map is covering. Furthermore,
if $t_{1} \ne t_{2}$ then ${\mathcal W}_{t_{1}\psi} 
\ne {\mathcal W}_{t_{2}\psi}$.
Hence, $\exp(t_{1} \psi) \ne \exp(t_{2} \psi)$.
Suppose that $\psi_{1}$ and $\psi_{2}$ are linearly
independent vectors, from the construction of $\exp$ it follows
that complex curves $\{\exp(t\psi_{1})|t \in {\nEW C} \}$ and 
$\{\exp(t\psi_{2})|t \in {\nEW C} \}$ intersect only at the
identity. These arguments imply that $\exp$ is a 
differentiable inclusion from $T_{id}{\mathcal T}$ to 
${\mathcal T}$. Since  $T_{id}{\mathcal T}$  is open and closed,
${\mathcal T}$ will be the image of $\exp$. 
From here it follows that the exponential
map is surjective.  
Finally, to show that the inverse map to the exponential map
is  differentiable, all we have to do is to remark that
\[ d \exp_{t v}= L_{\sigma(t)} : T_{id}{\mathcal T} \rightarrow T_{\exp(tv)}
{\mathcal T} \mbox{,} \]
where $L_{\sigma(t)}$ is the parallel transport along the 
geodesic $\sigma(t)$ connecting the $id$ to the  point $\exp(tv)$.
Obviously this map is invertible, and hence because of the inverse 
function theorem we have an isomorphism between the tangent spaces.
\end{proof}

\medskip\noindent
{\bf Remarks.} \ 
\begin{enumerate}
\item 
The totally geodesic manifold 
$D_{\psi} = \{\gamma_{\psi}(s) | s \in {\nEW C} \}$
is isometric to the disc $D_{R}$ of radius $R=|4/K_{\psi}|$ endowed with the
Poincar\'e metric
\[ \frac{4 R^{4} du^2}{(R^2-|u|^{2})^{2}} \mbox{.} \]
\item
The negativity of the curvature shows the instability 
of the geo\-desic flow in the sense of Arnold \cite{arnold}.
\item
The analogue of the Cartan-Hadamard result shown in Theorem~\ref{cartan} 
allows us to conclude that the $\mbox{\rm Diff}^{+}(S^{1})$ is homotopically
equivalent to $S^{1}$.
\item 
The infinite-dimensional Siegel disc is by definition
the set of Hilbert-Schmidt operators $T: H_{+} \rightarrow H_{-}$ 
such that $\det(I-TT^{\ast}) > 0$. As a consequence of our results,
it follows that the space ${\mathcal T}$ can be isometrically 
identified  with the infinite-dimensi\-onal Siegel disc. Indeed,
any two points in both spaces can be joined by a unique geodesic and
we have just shown that ${\mathcal T}$ is geodesically complete.

\end{enumerate}

\label{sec6}				
\section{Construction of Solutions to KdV from 
the Beltrami Equation}

In this section we will describe a procedure
to construct solutions to the periodic KdV 
equation.  The construction will be based on the 
following theorem:

\begin{thr}			
Let $\mu\in T_{id}{\mathcal T}^\ast$ be such that 
\[  \mu =
\begin{cases}
-\frac{1}{2} (1-|z|^2)^{2}\varphi(\bar{z}), & \quad  z\in \Di_{0} \\
 0,  & \quad  z \in \widehat{\nEW C}\setminus \Di_{0}
\end{cases}
\]
and $\|\mu \|_{\infty} < 1$. 
For $s$ in a neighborhood of $0$ 
take $f_{\infty,s}$ and $f_{0,s}$  as in the construction
of  Section~\ref{sec4}.

Then, $\exp_{\mu}(s)  = f_{0,s}^{-1} \circ f_{\infty,s}$ 
extends analytically as a totally geodesic two-real-dimensional
manifold in ${\mathcal T}$. 
\end{thr}

\proof
The idea is to show that in a neighborhood of
$0$, there exists a 1-1
correspondence  between $\exp_{\mu}(s)$ and the
totally geodesic submanifold constructed in Section~\ref{sec6}.
We recall the embedding of ${\mathcal T}$ into the Grassmannian given by  
\[{\mathcal T} \ni \sigma_{s} \mapsto {\mathcal W}_{s} \in Gr \mbox{,} \]
where 
\[{\mathcal W}_{s}\bydef \overline{ (\ker \bar{\partial}_{s\mu})
|_{S^{1}} } \mbox{.} \]
In order to complete the proof of the theorem we need the following: 

\begin{lm}			
For $s$ sufficiently small, there exists a linear operator 
$L_{\mu}:H_{+}\rightarrow H_{-}$ such that
\[{\mathcal W}_{s}=\mbox{\rm Graph}(sL_{\mu})\mbox{.} \]
\label{graph}
\end{lm}

\begin{proof} 
We denote by $\Gamma(\Di_{0},\Omega^{1,0})$
the set of smooth $(1,0)$-one-forms on $\Di_{0}$ of the form
$f dz$ such that $f$ extends to the circle,  
and mutatis mutandis  
$\Gamma(\Di_{0},\Omega^{0,1})$ the set of $(0,1)$ one-forms.
For sufficiently small $s$ let 
\[ s L_{\mu}:\Gamma(\Di_{0},\Omega^{1,0})\rightarrow
\Gamma(\Di_{0},\Omega^{0,1}) \mbox{,} \]
be a family of linear operators 
defined  by
\[ dz \mapsto s \mu_{\varphi} d \bar{z} \mbox{.} \]
Let 
\[ W_{\Di_{0},s\mu} \bydef \mbox{\rm Graph}(s L_{\mu}) 
\subset \Gamma(\Di_{0},\Omega^{1,0})\oplus  \Gamma(\Di_{0},\Omega^{0,1}) 
\mbox{.} \]

We will show that
${\mathcal W}_{s} = \overline{ W_{\Di_{0},s\mu} |_{S^{1}}\mbox{.}}$
Note that 
$f$ is a solution of
\[\dbar_{s\mu} f=0 \]
iff 
\[df \in W_{\Di_{0},s\mu}=\mbox{\rm Graph}(s L_{\mu}) \mbox{.} \]
This follows directly from the following computation:
\begin{eqnarray*}
df & = & \partial f dz + \dbar f d \bar{z} \\
   & = & \partial f \left(dz + \frac{\dbar f}{\partial f} d \bar{z}\right) 
\end{eqnarray*}
So, 
$df \in  W_{\Di_{0},s\mu}$ iff 
\[  \frac{\dbar f}{\partial f} = s  \mu \mbox{.} \]
From here the lemma follows just by restricting to the 
points of $S^{1}$ and using that the solutions of
$\dbar_{s\mu} f=0$ are of the form 
$f(z) = G(w(z))$ with $G$ analytic and $w$ an $s\mu$-quasiconformal
homeomorphism of the plane on itself.
\end{proof}

From Lemma~\ref{graph} it follows that there exists a 1-1 
correspondence  between $\exp_{\mu}(s)$ and the
totally geodesic submanifold constructed in Section~\ref{sec6},
namely $\exp(sL_{\mu})$. 

We are now ready to state a method for the construction of the solutions 
to the KdV equation starting with initial data of the form~(\ref{condit})

We summarize this discussion in the following procedure of constructing
periodic solutions of KdV:
\begin{enumerate}
\item
Given $u_{0}$ as in equation~(\ref{condit})
take  
\[ \phi(\exp{ix}) = \sum_{n \ge 0} a_{n} \exp(inx) \]
and construct
$s\mu_{\phi} = s(1-|z|^{2})^{2}\phi(\bar{z})$ for $z\in \Di_{0}$
and $s\mu_{\phi}=0$ outside the disk $\Di_{0}$.

\item
Let $\omega_{s\phi}$ be the solution of the Beltrami
equation
\[
\left\{ \begin{array}{l}
\dbar \omega - s \mu_{\phi}(z)  \partial \omega = 0  \mbox{,} \\
\omega(\rho) = \rho \mbox{ for   } \rho \in \{ -1,-i,1\}.
\end{array}
\right.
\]

\item
We define $f_{\infty,s}(z)=\omega_{s\phi}(z)$ for $z\in \Di_{\infty}$.
Set $f_{0,s}$ as in equation~(\ref{rie0}).

\item
Let for $z\in S^{1}$
\[ g(t,z)= f^{-1}_{0,s(t)}\circ f_{\infty,s(t)}(z) \mbox{.} \] 
\item
Choose a parameterization $s=s(t)$ so that
\[ \left\| \frac{d}{dt} g \right\|_{WP} = 1 \mbox{,} \]
where the $\| \ \|_{WP}$ is the norm w.r.t. the Weil-Petersson metric, i.e.,
the right-invariant metric defining the Riemannian structure of ${\mathcal T}$.

\item
If we take 
\[ u(t,x) = {\mathcal A} \left( 
\frac{\dot{g}(t,\exp(ix))}{g'(t,\exp(ix))} \right) \mbox{,} \]
where $'$ denotes the derivative w.r.t. $z$, $\cdot$ the time derivative
and ${\mathcal A}$ is the identification used in Corollary~\ref{andrey}, 
then $u$ satisfies the KdV equation by the Corollary~\ref{arn} in the
Appendix~1.
\end{enumerate}

\label{sec7}		
\section{Final Remarks}
\label{sec8}

\begin{enumerate}
\item 
As this manuscript was being finalized we became aware of a 
beautiful paper by Nag and Sullivan \cite{nagsul} where
the diffeomorphisms of the circle modulo rotations are endowed with
a Hilbert manifold structure modeled on the Sobolev space ${\mathcal H}^{1/2}$.
We emphasize the difference that in this paper we are concerned with
the Hilbert manifold structure of $\Dtp$. Hence our manifold is modeled
upon ${\mathcal H}^{3/2}$. 

\item
The idea of the curvature computation is closely related to the 
curvature computation of the Weil-Petersson metric on the 
moduli space of Calabi-Yau manifolds \cite{todo1}.
Other formulae for the curvature of some of the orbits of the
Bott-Virasoro group with a natural K\"{a}hler metric were obtained
by Kirillov and Yurev \cite{kirillov1,kirillov2}.
We expect in  a future work  to obtain explicit formulae for the 
curvature in terms of 
the Green function for the Laplace operator on the disk with the
Poincar\'e metric (following some ideas of Siu).

\item
The relation between metrics  of constant negative curvature 
and completely integrable systems is also present in the work of 
\linebreak
S.S.~Chern and K.~Tenenblat (see \cite{keti}).

\item 
A number of deep analytic results concerning existence and un\-iqueness 
for the periodic KdV in Sobolev space were recently obtained by 
J. Bourgain \cite{bour}. We remark that even the problem of local
existence for periodic KdV in Sobolev spaces is a non-trivial one
when $s\le 3/2$. See \cite{ponce}. Our method gives long time
existence for the solutions of KdV. We remark that our techniques
are geometric ones and 
can be applied to other orbits of the Bott-Virasoro group. We are currently 
investigating the relation between existence results for $L^{2}$ initial
data and the co-adjoint orbit of the Bott-Virasoro group isomorphic to
$\mbox{\rm Diff}(S^{1})$. (The classification of the co-adjoint
orbits of the Bott-Virasoro group was obtained in a result
of Kuiper's \cite{kuiper}.)

\item 
The relation between the KdV (or more generally the KP) hierarchy and
the Grassmannian is a standard fact from the theory of solitons
\cite{JimboMiwa, SegalW, McKeanS}. 
We remark, however, that the relation between the KdV equation and 
the Grassmannian in the present work 
is totally distinct from the one in \cite{SegalW}.

\item 
The  study of the
periodic KdV equation is naturally associated to Hill's operator and to
the theory of theta functions (see \cite{Novikov} and references therein).
It would be very interesting to connect the present results with
specific known examples, in particular with the
results in \cite{McKeanTrub1, McKeanTrub2, McKeanC, IsMcKeanTrub1}. 
We are currently working on such examples.
\end{enumerate}

\appendix
\section{Appendix 1: Arnold's point of view on Euler equations}

In the first part of this appendix we describe Arnold's ideas about the 
connection
between Euler equation and geodesic flows on finite dimensional
Lie groups. On the second part we describe the infinite dimensional
analogue between Euler's equation and geodesic flows due to Arnold.
For an introductory exposition to the material of this appendix we 
refer the reader to \cite{mratiu}.

\begin{itemize}
\item
{\it Arnold's approach {\em \cite{arnold}} to Euler's Equation in a Finite 
Dimensional Space}: 
Let $G$  be a finite dimensional Lie group, and let ${\mathcal G}$ be its
Lie Algebra. As usual we denote by
$ L_{g}h=g h$ the left multiplication (resp. $R_{g}h=h g$ the
right-multiplication), and by $L_{g \ast}$ the derivative of 
this transformation $ L_{g}$ (respec. $R_{g \ast}$). 
Let 
\[A:{\mathcal G} \rightarrow {\mathcal G}^{\ast} 
= {\rm Hom}({\mathcal G},{\nEW R})\] 
be a positive symmetric operator. By positive symmetric operator we mean  
\[ \langle A u | u \rangle  \  >  \    0  \ \  \ 
\mbox {    } \forall u \ne 0 \]
and \[ \langle A v |u\rangle \  = \ \langle A u |v\rangle  \mbox{,} \]
where $\langle  |\rangle $ denotes the duality bracket between
${\mathcal G}^\ast$  and ${\mathcal G}$, i.e.,
$ \langle F | \xi \rangle  = F(\xi) $ for $ F \in {\mathcal G}^{\ast}$  
and $ \xi \in {\mathcal G}$.

The operator $A$ defines a right-invariant metric on $G$.
Let $g(t)$ be a geodesic w.r.t. to this right-invariant metric defined
by $A$.
From the right-invariance of the metric, we have that
\[  R_{g^{-1} \ast} \dot{g}(t) = \dot{g}(0) = u_{0}  \mbox{,} \]
which is equivalent to 
\[ R_{g(t)\ast} u_{0} = \dot{g}(t) \mbox{.} \]

Let $u(t)= L_{g^{-1}(t) \ast } \dot{g}(t) 
=  L_{g^{-1}(t) \ast} R_{g(t) \ast} u_{0} $.
Since, $A:{\mathcal G} \rightarrow {\mathcal G}^{\ast}$ is an
isomorphism we will look at ${\mathcal G}$ as identified with 
${\mathcal G}^{\ast}$ by choosing an orthonormal basis 
in ${\mathcal G}$. 
The vector $m=Au_{0}$ is called the generalized angular momentum and
$M(t) = A u(t)$ is called the relative angular momentum.
We have the conservation law
\[ \frac{dm}{dt} = 0 \mbox{ ,} \]
meaning that $m$ is a constant of motion.
From here one obtains (see Section~D of Appendix~2, 
\cite{arnold}) Arnold-Euler's equation for 
$M(t)$, namely:
\[\frac{dM}{dt} = \{ u, M \} \mbox{,} \]
where $\{ , \}$ is the infinitesimal co-adjoint action.
We recall that the infinitesimal co-adjoint action
for $M\in{\mathcal G}^{\ast}$ and $ \xi,  \eta \in {\mathcal G}$ is given by:
\[ \{\xi,M\}(\eta) \bydef (\mbox{\rm ad}^{\ast}_{\xi} M )(\eta)  
=  \langle M | [ \xi , \eta ]  \rangle  \]

\item
{\it Arnold's  approach to Euler's Equation in an Infinite Dimensional Space}: 
In this case we restrict ourselves to the Bott-Virasoro group $\Gat$.
Bott proved that  (see \cite{bott}): 
\end{itemize}

\begin{thr}				
There exists  a group $\Gat$ that is a central extension  of 
$G=\mbox{\rm Diff}_{+}(S^{1})$ given by Bott's co-cycle 
$B: \mbox{\rm Diff}_{+}(S^{1})\times  \mbox{\rm Diff}_{+}(S^{1}) \rightarrow {\nEW R}$, where 
\[ B(\sigma_{1},\sigma_{2}) = 
\int_{S^{1}}
\log(\sigma_{1}(\sigma_{2})) ' \  d(\log \sigma_{2})'
\]
\end{thr}

\medskip\noindent
{\bf Remark.} \ 
The central extension $\Gat$ above is called the Bott-Virasoro
group.

\begin{defi}\label{vir}				
{\em 
The Virasoro Lie Algebra is a vector space $\gat={\nEW C} \oplus
{\mathcal G} $ where ${\mathcal G}$ is the Lie-algebra of complex
vectors-fields on $S^{1}$ and the Lie-bracket is defined by
\[
\left[(\lambda_{1},f\pdx),(\lambda_{2}, g \pdx )\right] =
(c_{0}(f,g) , \left[f\pdx,g\pdx\right]) \]
where
\begin{equation}
 c_{0}(f,g)=\frac{1}{2 \pi i } \int_{S^1 } f'''g \ dz 
\label{cocycle}
\end{equation}
}
\end{defi}

Note that this co-cycle extends as a {\em bounded skew-symmetric form}
on $\HSob^{3/2}(S^{1})$. Indeed, if $f\in \HSob^{3/2}(S^{1})$, then
$f''' \in \HSob^{-3/2}(S^{1})=$  \linebreak
$(\HSob^{3/2}(S^{1}))'$.
We remark that the tangent space at the identity to $\Gat$ 
is $\gat$.
The construction of Arnold-Euler's equation in the finite-dimensional
case leads us to the following natural definition in the
infinite-dimensional context: 

\begin{defi}				
{\em
Let $A$ be a positive operator from
$\gat$ to $\gat^{\ast}$, and suppose that 
$M(t)$ is a curve in $\gat^{\ast}$ satisfying
\begin{equation}
 \frac{d M}{d t} = \{ A^{-1} M(t), M(t)\} \mbox{,} 
\label{eq:euler}
\end{equation}
where the infinitesimal co-adjoint action is
given by:
\[ \{\xi,M\}(\eta)=\langle M | [\xi,\eta] \rangle  \mbox{.}  \]
Then, equation~(\ref{eq:euler}) will be called
the Arnold-Euler equation.
}
\end{defi}

The following result is well known  (see \cite{ovs,segal,segal2}) for
the special coadjoint orbit of the Bott-Virasoro group which is  
isomorphic to $\Dinfty$.

\begin{thr}				
The periodic KdV equation
\[ u_{t} = u_{xxx} + 6 u u_{x} \]
coincides with the Arnold-Euler equation on the K\"ahler co-adjoint orbits
of the Bott group.
\end{thr}

\proof
First we will compute the co-adjoint action 
of the Bott-Virasoro group on $\gat^{\ast}$ following  
Kirillov (see \cite{kirillov}). For that we use the following:   

\begin{thr}[Kirillov]					
The dual Lie algebra $\gat^{\ast}$ as a $\Gat$-module 
is isomorphic to the $\Gat$-module 
 \[ Z^{2}({\mathcal G}) \bydef \{c:{\mathcal G}\times
{\mathcal G} \rightarrow {\nEW R} \; |\;
c([\xi,\eta],\zeta)+c([\eta,\zeta],\xi)+c([\zeta,\xi],\eta)  = 0 \} 
\]
\end{thr}

The idea of the proof of Kirillov's result is based on two facts.
The first is the Gelfand-Fuchs theorem, which states that
\begin{multline*}
Z^{2}({\mathcal G}) =  \{ \lambda c_{0} + \delta  \ | 
\lambda \in {\nEW R},\; \text{  $c_{0}$ as in  eq. (\ref{cocycle}), and }\\
\delta[\xi,\eta] \bydef \langle \delta |\left[\xi,\eta\right]\rangle  
\mbox{ for } \delta \in {\mathcal G}^{\ast} \}.
 \end{multline*}

The second idea is the computation of the action 
of $\Gat$ on $Z^{2}({\mathcal G})$. Let
\[q \in {\mathcal G}^{\ast} = \{ q(x) \ dx^{\otimes 2} \} \mbox{.} \]
Kirillov proved that
\begin{equation}
\mbox{\rm Ad}^{\ast}(\lambda,\varphi)(t,q) = 
(t, \mbox{\rm Ad}^{\ast}( \varphi )
(q) + t h(\varphi )) \mbox{,}  
\label{eq:kir1}
\end{equation}
where
\[h(\varphi) =  {\mathcal S}(\varphi)\circ\varphi^{-1} \mbox{ }\]
and for $\eta \in {\mathcal G}$ 
\[ \langle \mbox{\rm Ad}^{\ast}( \varphi ) q | \eta \rangle =  
 \langle  q | \eta\circ\varphi^{-1} \rangle \mbox{.} \]

Here, ${\mathcal S}$  denotes the Schwarzian derivative 
defined in Section~\ref{sec2} (but here taken with respect to $x$).
The proof of Kirillov's theorem is a direct consequence of these
two remarks.

By a distribution of weight $\alpha$ we
mean a tensor field $f(z)(dz)^{\alpha}$ that changes according to the
standard rules, i.e.,
\[ f(z) (dz)^{\alpha} = g(w) (dw)^{\alpha} \]
for a change of coordinate $w=w(z)$. 
We now recall the following result of Lazutkin \cite{lazutkin} and Kirillov's.

\begin{lm}[Lazutkin, Kirillov]				
The set 
$Z^{2}({\mathcal G}) \equiv \gat^{\ast}$ can be identified with the 
${\mathcal G}$-module of Hill's operators
\[ \{ \lambda \pdx^{2} + q  \}  \]
acting on distributions of weight $-1/2$
as $\Gat$-modules. 
\end{lm}

The idea of the proof is based on the following fact:
Suppose that $\eta_{1}$ and $\eta_{2}$ are two linearly independent solutions
of Hill's equation 
\[ (\lambda \pdx^{2} + q  )\eta = 0 \mbox{,} \]
then 
\[ q= {\mathcal S}\left(\frac{\eta_{1}}{\eta_{2}}\right) \mbox{.} \]

If we perform a  change of variables $x \mapsto \sigma^{-1}(x)$ and
use the property of the Schwarzian derivative
\[{\mathcal S}(\varphi\circ\psi) 
= ({\mathcal S}(\varphi)\circ\psi)(\psi_{x})^{2} 
+ {\mathcal S}(\psi) \mbox{,} \]
we get that $\eta_{i}\circ\sigma^{-1}(x)$
will satisfy the Hill's equation
\[ (\lambda \pdx^{2} + \widetilde{q}  )\eta = 0 \mbox{,} \]
with
\begin{eqnarray*}
\widetilde{q}(x) & = &  
{\mathcal S}\left(\frac{\eta_{1}(\sigma^{-1}(x))}{\eta_{2}
(\sigma^{-1}(x))}\right) \\
 & = & q\circ \sigma^{-1}(x)((\sigma^{-1}(x))_{x})^{2} 
+{\mathcal S}(\sigma^{-1}(x))  \mbox{.}
\end{eqnarray*}
From here and the Kirillov formula (\ref{eq:kir1}) the lemma follows.

\begin{itemize}
\item
{\it Computation of the co-adjoint action of $\xi\in {\mathcal G}$
on $\gat^{\ast}$}:

We start by noticing, from the discussion above, that
\[\gat^{\ast} \cong \{ (\lambda, q dx^{\otimes 2})  \} \mbox{ .} \]
The computation will be done in three steps.

\begin{enumerate}
\item
Computation of the infinitesimal action of $\xi$
on quadratic differentials.
We proceed in the standard way. Let $g_{t}(x)$ be a one-parameter
family of diffeomorphisms of the circle and  such that 
\[ g_{t}(x) = x + t \xi + \mbox{\rm h.o.t.}  \]
We will compute the derivative $\delta b$ of the
family of quadratic differentials
$b(g_{t}(x))(d(g_{t}(x)))^{\otimes 2}$ at $t=0$.
Hence,
\begin{align*}
& b(g_{t}(x))(d(g_{t}(x)))^{\otimes 2} \\
& = b(x+ t \xi + \mbox{ h.o.t.}) (d(x+t\xi+\dots))^{\otimes 2}  \\
& =  (b(x) + t b_{x}(x) \xi + \mbox{ h.o.t})
(dx+t\xi_{x}dx + \mbox{ h.o.t.})^{\otimes 2} \\
& =  b(x)(dx)^{\otimes 2} + t (b_{x}\xi+2\xi_{x}b)(dx)^{\otimes 2} 
+  \mbox{ h.o.t.} 
\end{align*}
Hence,
\[ \delta b = (b_{x}\xi + 2 \xi_{x} b)(dx)^{\otimes 2} \mbox{.} \]

\item
Computation of the pairing of $\gat$ and $\gat^{\ast}$.
We start by recalling that the dual $\gat^{\ast}$ can be
identified to the set
\[ \{(t,b)| t\in {\nEW R} \mbox{ and $b$ is a quadratic differential} \} 
\mbox{.}\]
Here, by quadratic differential, we mean an element of
the form $b(x)dx^{\otimes 2}$, where $b$  for a local coordinate-chart is
an element of $\HSob^{-3/2}$.
So that the pairing between a point $(\mu,\eta) = (\mu, q d/dx) \in \gat$ and
a point $(s,b (dx)^{\otimes 2})$ is given by
\[ s\mu + \int_{S^{1}} b q dx \mbox{.} \]
This last integral being performed in the sense of the
pairing between $\HSob^{3/2}$ and $\HSob^{-3/2}$.
We also notice that this is independent of the parametrization
of the circle.
Now we compute
\begin{align*}
&\langle (t,b) |\left[(\lambda,\xi),(\mu,\eta)\right] \rangle  \\
&=\left\langle (t,b) \Big|\left(\frac{1}{2\pi i}
\int_{S^1}\xi_{xxx}\eta dx,[\xi,\eta]\right)\right\rangle \\
&=\frac{t}{2\pi i}\int_{S^{1}}\xi_{xxx}\eta dx 
+  \frac{1}{2\pi i}\int_{S^{1}} [\xi,\eta]b dx .
\end{align*}

\item 
Computation of the infinitesimal co-adjoint action of $\xi$
on $\gat^{\ast}$.
We will combine  the two items above to compute
\[(t,\widetilde{b}) \bydef \mbox{\rm ad}^{\ast}_{(\lambda,\xi)} (t,b) 
\mbox{.} \]
Hence,
\[(t,\widetilde{b})(\mu,\eta)=
\langle \  (t,b)  \ | \ [(\lambda,\xi),(\mu,\eta)] \ \rangle  \mbox{.} \]
From  the above results we get
\begin{equation}
 \widetilde{b}= \frac{1}{2}\lambda \xi_{xxx}+2\xi_{x}b+\xi b_{x} 
\mbox{.} 
\label{eq:coad}
\end{equation}
The crucial point now is that for $b\in \HSob^{3/2}$ the mapping
\[ \xi \mapsto \widetilde{b} \]
defined in equation~(\ref{eq:coad}) is a linear (unbounded)
operator from the dense subset $\HSob^{+3/2}$ of $\HSob^{-3/2}$
into $\HSob^{-3/2}.$
\end{enumerate}

\item
{\it Computation of the Arnold-Euler equation}:

We now define $A:\gat \rightarrow \gat^{\ast}$
a positive symmetric operator, 
with the help of Kirillov's form.
It is defined at  a given point $u\in  \gat^{\ast}$
by
\[ \langle A \xi | \eta \rangle = \Omega ({\mathcal I} \xi , \eta) \mbox{,} \]
where ${\mathcal I}$ is the complex structure, and Kirillov's
formula is given by
\[ \Omega(\xi,\eta)(u) = \langle u | [\xi,\eta] \rangle \mbox{,} \]
for  $\xi$ and $\eta$ in $\gat$.
We remark that all these objects extend by right translation to the
whole orbit. Furthermore, 
\[ \Omega ({\mathcal I} \xi , \xi) \ge 0 \]
and 
\[ \Omega ({\mathcal I} \xi , \eta) 
= \Omega ({\mathcal I} \eta , \xi) \mbox{.} \]
Finally, the above defined symmetric form realizes the
Weil-Peter\-sson metric. 
At this point we choose a parametrization of the circle, and
with the help of such parametrization 
$x:{\nEW R} \rightarrow S^{1}$ we have an orthonormal basis given
by 
\[ e_{n}(x) = \frac{z^{n-2}}{\sqrt{n^{3}-n}}, \quad n \ge 2 \mbox{, } \]
where $z=\exp(ix)$.

We remark that in a local system of coordinates, the function
$q$ above belongs to $\HSob^{3/2}$ and the second component of the map
$A^{-1}$ has image 
in  $\HSob^{-3/2}(\Rin,\Rin)$.
Now we use equation~(\ref{eq:coad})   
where $\lambda=1$, $\xi=q d/dx$ and $b=qdx^{\otimes 2}$. 
This yields in the sense of $\HSob^{-3/2}$   (modulo linear changes of
variables)
\begin{equation}
\frac{dq}{dt}=\left(\frac{1}{2} q_{xxx}+3qq_{x}\right) \mbox{ .}
\label{segaln}
\end{equation}
Hence, the Arnold-Euler equation gives the KdV 
equation.
The Arnold-Euler 
equation (as remarked by Ovsienko and Khesin in \cite{ovs}) is the 
Hamiltonian flow on the symplectic co-adjoint orbits
in $\gat^{\ast}$ generated by the Hamiltonian function
\[ q \mapsto \frac{1}{2} 
\langle  (1,q (dx)^{\otimes 2}) |A^{-1}(1,q (dx)^{\otimes 2}) \rangle  
\mbox{.} \]

\end{itemize}

According to Kirillov (see also \cite{witten2}) our manifold ${\mathcal T}$
is a co-adjoint orbit of the Bott group. 
\endproof

Repeating the argument of		
Arnold as presented on the first part of this Appendix we conclude:

\begin{thr} \label{geodflow}
The geodesic flow on the K\"{a}hler manifold ${\mathcal T}$ with respect to 
the right invariant K\"{a}hler metric defined in equation~(\ref{kahler})
induces the KdV flow on $T_{id}{\mathcal T}$
by means of left translations
to the identity. 
\end{thr}

\begin{proof}
Let $g(t)=g(t;\cdot)$ be a geodesic w.r.t. the K\"{a}hler metric on
${\mathcal T}$. 
From the right-invariance of the K\"{a}hler metric on ${\mathcal T}$
we have that 
\[  R_{g^{-1} \ast} \dot{g}(t) = \dot{g}(0) = u_{0}  \mbox{.} \]
As before, this is equivalent to
\[ R_{g(t)} u_{0} = \dot{g}(t) \mbox{.} \]
Let $u(t,x)= L_{g^{-1} \ast } \dot{g}(t) $.
By choosing an o.n. basis we will identify the 
tangent and the cotangent spaces at the identity of ${\mathcal T}$.
Let $A:T_{id} {\mathcal T} \rightarrow T_{id} {\mathcal T}^{\ast}$ 
be this identification.
Using the result of Arnold concerning the geodesic flow, proved in Appendix~2 
of Reference~\cite{arnold}, we conclude that
\[ \frac{d A u(t,x)}{dt} = \{ u(t,x) , A u(t,x) \} \mbox{.} \]
From  formula~(\ref{segaln}), it follows that this last
equation is equivalent to KdV. 
\end{proof}

\begin{co}				
\label{arn}
Let $g(t,\exp(ix)) = \exp(i\phi(t,x))$ be a geodesic in 
${\mathcal T}$ with initial velocity $\dot{\phi}(0,x)$ where 
\[ A \dot{\phi}(0,x) = 
\sum a_{n} e^{inx} \in 
\HSob^{3/2}({\nEW R};{\nEW R}) \mbox{.} \]

Then,
\[ q(t,x) = A \left( \frac{\dot{\phi}(t,x)}{\phi_{x}(t,x)}
\right)   \mbox{, } \]
is a solution of the KdV equation in $\HSob^{3/2}$.
\end{co}

\begin{proof}
We have shown that the geodesic flow along ${\mathcal T}$
extends for all time in the holomorphic directions.
Since the geodesic $g$ extends for all time  it follows 
that so does the velocity vector $\dot{g}$. 
Hence,
$ q(t,x) = A( \dot{\phi}(t,\exp(ix))/\phi_{x}(t,\exp(ix)) )$ 
is in $\HSob^{3/2}$ for all time and satisfies the KdV equation
because of Theorem~\ref{geodflow}.
\end{proof}

\section{Appendix 2: Review of Some Basic\\ Facts of Complex Analysis}
\label{secapp}

\subsection{Introduction}

In this Appendix we shall summarize some well known facts 
from the Theory of Teichm\"uller spaces. The purpose is two-fold.
First, we hope that this will make the results of the paper accessible
to a wider audience. Second, it will make the notation and the 
definitions used easily available. We stress, however, that this appendix
is not intended to be a survey of the theory. For that please refer to
\cite{Ahlfors, lehto, lehtovirt, gardiner, pomme, nag}.

In what follows we shall denote by ${\nEW H}$ the upper-half-plane 
and by $\widehat{\nEW R}$ the one point compactification of the real line.

\subsection{Quasiconformal Mappings}

A {\em quasiconformal mapping} 
$f:{\mathcal D} \rightarrow f({\mathcal D}) \subset {\nEW C}$ 
is a homeomorphism that is absolutely continuous on lines, the partial
derivatives are locally square-integrable, and $f$
satisfies a.e. the  Beltrami equation
\[ \dbar f - \mu \partial f = 0 \mbox{,} \]
for some measurable function $\mu$ such that
\[ k \bydef \sup_{z\in{\mathcal D}} |\mu(z)| < 1 \mbox{.} \]
We shall say that $f$ is $K$-quasiconformal if
\[ K = \frac{1+k}{1-k} \mbox{.} \]
The complex valued function $\mu$ is called the {\em complex dilation}
of the quasiconformal mapping $f$. 

We introduce the Banach space $B_{p}$ of complex functions $w(z)$, 
$z\in {\nEW C}$ which vanish at $z=0$ satisfy a H\"older condition
with exponent $1-(2/p)$ and have generalized derivatives in $L_{p}$.
The norm in $B_{p}$ is given by:
\[ \|w\|_{B_{p}} = H_{1-2/p}[w] + \|\partial w\|_{L^{p}} +
\|\dbar w\|_{L^{p}} \mbox{,} \]
where 
\[H_{1-2/p}[w] \bydef \sup_{z_{1}\ne z_{2}} 
\frac{|w(z_{1})-w(z_{2})|}{|z_{1}-z_{2}|^{1-2/p}} \mbox{.} \]

The existence of a quasiconformal homeomorphisms for a given
measurable function $\mu$  in a region $\Omega \subset {\nEW C}$
is consequence of the following result: (See \cite{bersjohn},  page 269.)

\begin{thr}			
Let $\mu \in L^{\infty}$ be such that $\|\mu\|_{L^{\infty}} < 1$.
Then, for every $p>2$ sufficiently close to $2$ the equation
\[ \dbar w - \mu \partial w = \sigma \]
has a unique solution in $B_{p}$.
\end{thr}

As a consequence of the result above one gets: \cite{bersjohn}

\begin{thr}\label{exist}				
Given $\mu$ measurable on the plane such that $\|\mu\|_{L^{\infty}} < 1$,
there exists a unique quasiconformal homeomorphism with 
complex dilation $\mu$ of the plane onto itself with fixed points
$0$, $1$ and $\infty$.
Furthermore, if $\mu$ vanishes outside a compact subset of ${\nEW C}$,
then  there exists a unique quasiconformal homeomorphism $f^{\mu}$  with  
complex dilation $\mu$  such that $ f^{\mu}(0)=0 $
and $\partial f^{\mu} - 1 \in L^{p}({\nEW C})$ for some $p$. 
\end{thr}

An analogous result for the disk $\Di_{0}$ is 

\begin{thr}				
Given $\mu$ measurable on the disk $\Di_{0}$ such $\|\mu\|_{L^{\infty}} < 1$,
there exists a unique quasiconformal homeomorphism with  
complex dilation $\mu$ of the disk onto itself  with fixed points 
$0$ and  $1$. 
\end{thr}

\subsection{Quasisymmetric Functions}

A quasisymmetric function on $S^{1}$ is a homeomorphism 
$h:S^{1}\rightarrow {\nEW C}$ for which there exists $M$
such that for all $z_{1}$, $z_{2}$, and $z_{3}$ in $S^{1}$
\[ |z_{1}-z_{2}|=|z_{2}-z_{3}| \Rightarrow |h(z_{1})-h(z_{2})| \le 
M |h(z_{2})-h(z_{3})| \mbox{.} \]

Quasisymmetric functions play an important role in Teichm\"uller theory 
\cite{Ahlfors}. We shall say that a quasisymmetric function 
$\phi:S^{1}\rightarrow S^{1}$ is 
{\em  normalized}  if it fixes the points $-1$, $-i$, and $1$. 

It is usual to consider quasisymmetric functions on the real line,
in which case an equivalent definition is given. 
A strictly increasing homeomorphism 
$h:\widehat{\nEW R} \rightarrow \widehat{\nEW R}$,
satisfying $h(\infty)=\infty$ is called {\em $\lambda$-quasisymmetric }
if for every $x\in {\nEW R}$ and $t > 0$ we have 
\begin{equation}
\label{double}
 \frac{1}{\lambda} \le \frac{h(x+t) - h(x)}{h(x) - h(x-t)} \le 
\lambda \mbox{.} 
\end{equation}

One of the most important results of the theory is that quasisymmetry of $h$
is the necessary and sufficient condition for $h$ to be a boundary function
of a quasiconformal self-mapping of the upper-half plane ${\nEW H}$
fixing $\infty$. This is the content of the next two results, 
the second of which is due to Beurling and Ahlfors. (See \cite{lehto} 
pages 31 and 33.)

\begin{thr}			
Let  $f$ be a K-quasiconformal mapping of ${\nEW H}$ onto itself such that
$f(\infty)=\infty$. Then, $f$ can be extended to a continuous function
$\widetilde{f}$ on the closure  
$\widehat{\nEW R}\cup {\nEW H}$ of ${\nEW H}$. The boundary function 
$h=\widetilde{f}|_{\widehat{\nEW R}}$ satisfies the inequalities in
equation~(\ref{double}). Furthermore, the dependence of  
$\lambda$ on $K$  in equation~(\ref{double}) is continuous and  
$\lambda(1)=1$.
\end{thr}

The converse of this result is given by 

\begin{thr}[Beurling-Ahlfors]			
\label{ahlbeu}
Let $h$ be $\lambda$-quasisymmetric, then for $y\ge 0$ and $x\in {\nEW R}$ 
\begin{align*}
f(x+iy) & \bydef \frac{1}{2}\int_{0}^{1} (h(x+ty)+h(x-ty)) \, dt \\
& \quad+ i \int_{0}^{1} (h(x+t y) - h(x-ty)) \, dt \mbox{, } 
\end{align*}
is a $K$-quasiconformal mapping of ${\nEW H}$ onto itself. 
Furthermore, it agrees with $h$ for $y=0$, it is continuous, 
and $f(\infty)=\infty$. The maximal dilation $K$ is bounded by $K(\lambda)$ 
that depends continuously on
$\lambda$ and tends to $1$ as $\lambda$ goes to $1$.
\end{thr}

We now remark: 
\begin{enumerate}
\item 
Let  $h_{1}$ and $h_{2}$ be quasisymmetric 
functions and $f_{1}$ and $f_{2}$ their quasiconformal Beurling-Ahlfors
extensions. Then, the quasisymmetry constant of $f_{2}\circ f_{1}^{-1}$
tends to $1$ if so does the quasisymmetry constant $\lambda$ 
of $h_{2}\circ h_{1}^{-1}$.
\item 
Suppose $h=h(\cdot,\xi)$ depends analytically on the parameter $\xi$ and
$\partial_{\xi} h(\cdot,\xi) $ is integrable. Then, 
the Beurling-Ahlfors extension given above also depends
analytically on this parameter.
\end{enumerate}

\subsection{The Sewing Problem}
A number of problems in complex analysis reduce to the following
{\em sewing problem}\cite{lehto,lehtovirt,kirillov1}:
Given $\phi$ a quasisymmetric homeomorphism  of the circle find a pair of
homeomorphisms $(f_{0},f_{\infty})$ such that
\begin{itemize}
\item[(a)]
$ f_{0} :  \overline{\Di}_{0} \longrightarrow  f_{0}(
\overline{\Di}_{0}) \subset \widehat{\nEW C}$ and
$f_{\infty} : \overline{\Di}_{\infty} \longrightarrow f_{\infty}( 
\overline{\Di}_{\infty}) \subset  \widehat{\nEW C} $,
where $f_{0}$ and $f_{\infty}$ are conformal in the interior of their
domains of definition,
\item[(b)] 
The sets $f_{0}( \Di_{0})$ and
$f_{\infty}( \Di_{\infty})$ are complementary Jordan domains,  and
\item[(c)] For every $z\in S^{1}$
\[\phi(z) = f_{0}^{-1} \circ  f_{\infty}(z) \mbox{.} \]
\end{itemize}

It is obvious that the problem, as stated above has many solutions if it has
one. Indeed, let $(f_{0},f_{\infty})$ be a solution of the problem. Then,
left composition by any linear fractional transformation  $T$  
is such  that $(T\circ f_{0},T\circ f_{\infty})$ is also a solution of the
sewing problem. In order to solve this problem of nonuniqueness we introduce
the following:

\begin{defi}				
{\em
We shall say that a homeomorphism $\phi:S^{1} \rightarrow S^{1}$ is
normalized if it fixes the points $-1$, $-i$, and $1$. 
We shall say that a solution $(f_{0},f_{\infty})$ of the sewing problem is 
normalized if both $f_{0}$ and $f_{\infty}$ fix $-1$, $-i$, and $1$.
}
\end{defi}

We remark that the choice of the points $-1$, $-i$, and $1$ is mere 
convenience. The sewing problem stated above is usually considered
in the context of the upper half plane ${\nEW H}$ and its boundary 
$\widehat{\nEW R}$. In that context, the normalization imposed consists
of fixing $0$, $1$ and $\infty$. Obviously, we are translating all such 
notions to the context of $\Di_{0}$ and $S^{1}$ by means of the
linear fractional transformation 
\[ z = \frac{\zeta -i}{\zeta+i} \mbox{.} \]

The following result settles the sewing  problem
\cite{pfluger, lehto, lehtovirt}: 

\begin{thr}\label{sewing}			
Let $\phi$ be a normalized quasisymmetric function. Then, the sewing problem
for $\phi$ has a unique normalized solution pair $(f_{0},f_{\infty})$.
\end{thr}

The main steps in the proof of this result are:
First, extend $\phi$ continuously to the interior of the disk as
a quasiconformal mapping $\widetilde{\phi}$ using the Beurling-Ahlfors result. 
Second,  let $\kappa[\widetilde{\phi}]$ be the complex dilation of
$\widetilde{\phi}$ and construct $F$ the unique  {\em normalized}
solution of the problem 
\begin{equation}
 \dbar F - \mu \partial F = 0 \mbox{,}
\label{dbarf}
\end{equation}
where
\begin{equation}
 \mu(z)  =  
\begin{cases}
\kappa[\widetilde{\phi}](z),  & \quad  z\in \Di_{0} \\
 0,  & \quad   z \in \widehat{\nEW C}\setminus \Di_{0}.
\end{cases}
\label{dbarmu}
\end{equation}
Third, let 
\[ f_{0} \bydef F |_{\overline{\Di}_{0}} 
\circ \widetilde{\phi}^{-1} \mbox{,}\]
and 
\[ f_{\infty} \bydef F|_{\overline{\Di}_{\infty}} \mbox{.}\]
It is easy to check that $f_{0}^{-1}\circ f_{\infty}(z) = \phi(z)$
for $z\in S^{1}$ and that both $f_{0}$ and $f_{\infty}$ are analytic
inside their domains of definition. The normalization of the solution
pair $(f_{0},f_{\infty})$ follows from that of $F$ and of
$\phi$.
Uniqueness (as well as the above argument) can be found in \cite{lehto}
page 101.

The next proposition was used in Section~\ref{sec2} to guarantee
that $\mu_{\phi}$ as in Definition~\ref{def1} corresponds to a
complex dilation of some quasiconformal mapping.

\begin{pr}				
Given $\epsilon>0$, there exists $\delta>0$ such that
if $\phi$ is $\lambda$-quasisymmetric with $|\lambda-1|<\delta$,
then 
\[ \|\mu_{\phi}\|_{\infty} = \sup_{z\in\Di_0}  (1-|z|^{2})^{2}
|{\mathcal S}[f_{\infty}](1/\bar{z})| < \epsilon \mbox{.} \]
\end{pr}

\begin{proof}
We follow the notation of Theorems \ref{ahlbeu} and \ref{sewing}.
The univalent conformal function $f_{\infty}$ is the restriction of a 
quasiconformal mapping $F$ of the extended plane  satisfying
equation~(\ref{dbarf}) 
with complex dilation given by (\ref{dbarmu}). It is known 
that\footnote{See for example Theorem~3.2 page 72 of \cite{lehto}.}
if $g$ is a quasiconformal mapping of the plane of complex dilation
$\widetilde{\mu}$ and 
such that $g$ restricted to the disc $\Di_{0}$ is conformal, then
\[\sup_{z \in \Di_{0}}(1-|z|^{2})^{2}  |{\mathcal S}[g](z)| 
\le 6 \|\widetilde{\mu}\|_{\infty}\mbox{.} \]
Writing $g(z)=F(1/z)$ and using properties of the Schwarzian
derivative, it follows that
\[ \sup_{z \in \Di_{0}} |(1-|z|^{2})^{2} {\mathcal S}[f_{\infty}](1/\bar{z})|
\le 6 \sup_{z\in \Di_{0}} |\kappa[\widetilde{\phi}](z)| \mbox{.} \]
Using Theorem~\ref{ahlbeu} we get that if $\phi$ is $\lambda$-quasisymmetric,
then
\[  \sup_{z\in \Di_{0}} |\kappa[\widetilde{\phi}](z)| \le 
\frac{K(\lambda)-1}{K(\lambda)+1} \mbox{,} \]
where $K(\lambda)\rightarrow 1$, when $\lambda \rightarrow 1$.
This proves that $\mu_{\phi}$ given by
\[
\mu_{\phi}(z) =
\begin{cases}
(1 - |z|^{2})^{2} {\mathcal S}(f_{\infty})(1/\bar{z}), & \quad 
z \in \Di_{0} \\
0, & \quad  z \in {\nEW C} \setminus \Di_{0}
\end{cases}
\]
 can be made arbitrarily small in the $L^{\infty}$ norm by taking $\phi$ 
a $\lambda$-quasi\-symmetric homeomorphism with $\lambda$ close to $1$.
\end{proof}

As a simple corollary of the above result we get

\begin{co}				
If $\phi$ is sufficiently close to the identity in the $C^{1}$ 
topology, then $\mu_{\phi}$ as in Definition~\ref{def1} has 
$\|\mu_{\phi}\|_{\infty} < 1 $. 
\end{co}

\subsection{The Result of Ahlfors and Weill}

The next result due to Ahlfors and Weill \cite{ahlweill}
plays an important role in the theory of Teichm\"{u}ller 
spaces. It is usually stated in the context of ${\nEW H}$ and
${\nEW H}^{\ast}$  \cite{gardiner}.
For convenience we are rewriting it in the context of $\Di_{0}$
and $\Di_{\infty}$.

\begin{thr}[Ahlfors-Weill]			
\label{aw}
Let $\varphi: \Di_{\infty} \rightarrow \widehat{\nEW C}$ be holomorphic
such that
\[  (1-|z|^{2})^{2} | \varphi(z) |  < 2  \mbox{ , } \forall z \in 
{\nEW D}_{\infty} \mbox{.} \]
Take, 
\[\mu(z)  =
\left\{ \begin{array}{cl}
-\frac{1}{2} \frac{(1-|z|^{2})^{2}}{\bar{z}^{4}} \varphi(1/\bar{z}),   
& \mbox{}  z\in \Di_{0} \\
 0,  & \mbox{}   z \in \Di_{\infty} \mbox{.}
\end{array}
\right.
 \]
Then, for any quasiconformal mapping $F$ on the Riemann sphere 
satisfying
\begin{equation}
(\dbar - \mu \partial ) F  = 0 \mbox{,}
\label{bel1}
\end{equation}
we have that 
\[ \varphi(z) = {\mathcal S}[F](z) \mbox{.} \]
\end{thr}

We shall not repeat the proof of this result here.
It could be found in page 100 of \cite{gardiner}. 
We shall, however, highlight the main points in
the proof. This, we believe, will give some insight
in the connection between Teichm\"uller's theory and the
Schr\"odinger equation.
The argument hinges upon the fact that the Schwarzian derivative
of ratios of independent solutions of the Schr\"odinger equation 
\begin{equation} 
y'' + \frac{1}{2} \varphi y = 0 \mbox{ ,}
\label{sch}
\end{equation} 
gives back the potential ${\varphi}$. More precisely, we have the following 

\begin{lm}\label{schw}			
Let $\varphi: {\mathcal D} \rightarrow {\nEW C}$ be
holomorphic. Set
\[ {\mathcal R}[\varphi] \bydef \left\{ w= \frac{y_{1}}{y_{2}} \; | \;
\text{ $\;y_{1}$ and $y_{2}\;$  l.i. solutions of (\ref{sch})  on
$\; {\mathcal D} $} \right\} \mbox{,} \]
and 
\[ {\mathcal S}^{-1}[\varphi] \bydef 
\{ w : {\mathcal D} \rightarrow {\nEW C} \mbox{ holomorphic} \; |
\; {\mathcal S}[w] = \varphi \} \mbox{.} \]
Then,
\[ {\mathcal R}[\varphi] = {\mathcal S}^{-1}[\varphi] \mbox{.} \]
\end{lm}

Now, the idea of the proof is based on the explicit construction
of a quasiconformal mapping satisfying (\ref{bel1}) (at least under
some simplifying assumptions), namely:
\[ F(z) = 
\left\{ \begin{array}{cl}
\frac{y_{1}(1/\bar{z}) + ( z - \frac{1}{\bar{z}} )y_{1}'(1/\bar{z})} 
{y_{2}(1/\bar{z}) + ( z - \frac{1}{\bar{z}} )y_{2}'(1/\bar{z})},
& \mbox{ }  z\in \Di_{0} \\
 \frac{y_{1}(z)}{y_{2}(z)},   
& \mbox{ }   z \in \widehat{\nEW C}\setminus \Di_{0} \mbox{,} 
\end{array}
\right. \]
where, $y_{1}$ and $y_{2}$ are solutions of equation (\ref{sch}). 
Once one establishes that $F$ as defined above is a solution of (\ref{bel1})
it follows from the Lemma~\ref{schw} that 
\[ \varphi={\mathcal S}[y_{1}/y_{2}] \mbox{.} \]
The result now follows using the invariance of the Schwarzian derivative 
given by the formula
\[{\mathcal S}(T\circ F) = {\mathcal S}(F) \mbox{,} \]
where  $T$ is any linear fractional
transformation, and the fact that different quasiconformal solutions
of (\ref{bel1}) are related by a linear fractional transformations. 

\medskip\noindent
{\bf Remark.} \ 
It is well known the important role played by the Schr\"odin\-ger
operator in the study of the KdV equation. It is also known the 
importance of the Riemann-Hilbert boundary value problem in the solution
of many completely integrable systems. 
The connection of the Schwarzian derivative with the Schr\"odinger
equation described above seems to indicate the fact that the Schr\"odinger
equation is entering the picture once more. Here, in the form of
an explicit way to produce solutions to the Beltrami equation.

\newcommand{\rn}{{\mathbb R}^n}
\newcommand{\m}{\medskip}

\subsection{ The Construction of the Normalized Solutions to the
Beltrami Equation}

In this part of the appendix we show how to construct a
 solution to the Beltrami equation with zero initial condition such 
that derivative of the solution minus certain monomials are in 
$L^p$ with $p\geq 2$.

{\m}
We recall that 
 if $f : A \rightarrow {\nEW C}$ has derivatives in $L^1(A)$ and $ D$ is a 
 domain such that $\bar{D} \subset A$,  then 
 \begin{equation}
 f(z) = \frac{1}{2 \pi i} \int_{\partial D} \frac{f(\zeta)}{\zeta-z}
dS(\zeta) - \frac{1}{\pi} \int\!\!\int_{D}
\frac{\bar{\partial}f(\zeta)}{\zeta - z}d\xi d\eta . \label{eq:3}  
 \end{equation}

\noindent This is the generalized Cauchy formula,  which is an immediate 
consequence of Green's formula. Hence, if $f(z)\rightarrow 0$, as 
$z \rightarrow \infty$, and if 
$f$ has derivatives in $L^1({\nEW C})$ it follows that 
\begin{equation}
f(z)=-\frac{1}{\pi} \int\!\!\int_{\nEW C} 
\frac{\bar{\partial}f(\zeta)}{\zeta - z}d\xi d\eta . \label{eq:4} 
\end{equation}
In other words, 
$f = {P} \bar{\partial} f$, where $P$ was defined in equation~(\ref{defP}).
With this information in hand we proceed to 
find a  solution for Beltrami's equation with the properties described above. 
More precisely,

\begin{thr} \label{theorem:b1} 			
For any integer $n\geq 1$ there exists 
at least  one solution $ \omega^{(n)}$ of
\begin{equation} 
\left\{ \begin{array}{l}
\bar{\partial}_{\mu_\phi} \omega^{(n)}  =   0  \\
\partial_{z}\omega^{(n)}- nz^{n-1} \in L^{p} 
\mbox{ \ {\rm for some}  $p  > 2 $}
\label{eq:5} \\
\int_{S^1} \omega^{(n)}  =   0.
\end{array} \right.
\end{equation}
\end{thr} 

\begin{proof} 
We first find a solution to 
\begin{equation}
\left\{ \begin{array}{l}
\bar{\partial}_{\mu_\phi} f^{(n)}  =  0 \label{eq:6}\\
\partial f^{(n)}- nz^{n-1} \in L^{p} \mbox{ for some $ p  > 2. $}
\end{array} \right.
\end{equation}

\noindent 
The solution is not unique. It will become unique when we 
impose the third condition in~(\ref{eq:5}). We will suppose 
$f(\infty) = \infty $. Then, in a neighborhood of $\infty$, we have 

\begin{equation}
 f^{(n)}(z) = g(z) + \sum_{i=0}^{\infty} b_{i} z^{-i}  
= z^{n} + G(z) + \sum_{i=0}^{\infty} b_{i} z^{-i},  \label{eq:7}
\end{equation}

\noindent where $g(z)$ and $G(z)$ are entire. 
This is a consequence of $\mu_{\phi} = 0 $ in ${\nEW C}\setminus \Di_{0}  $. 
Since $\partial_{z} f^{(n)}- nz^{n-1} \in L^p $ for some $p > 2$,  
we have $G(z) = A_{0} $,  with $ A_{0}$ a constant to be determined
below. Thus,
$$ f^{(n)}(z) = z^{n} + A_{0} + \sum_{i=0}^{\infty} b_{i} z^{-i}, $$

\noindent  
Since $f^{(n)} $ satisfies the Beltrami equation, its derivatives are
in $L^2$ locally (\cite{lehto}) and hence the derivatives of $f^{(n)}
- z^{n} -A_{0} \in L^2({\nEW C})$.  Note that $L^{2}(K) \subset
L^{1}(K)$ for any $K$ compact and by~(\ref{eq:7}) in a neighborhood of
$\infty $ it follows that $ f^{(n)} -z^{n}- A_{0} $ has derivatives in
$L^1$.  Hence their derivatives are in $L^{1}({\nEW C})$ and applying
the Cauchy generalized formula it follows that
$$ F = f^{(n)} - z^{n}- A_{0} = {P}\bar{\partial}f^{(n)} =
{P}\bar{\partial} F  \mbox{.} $$ 
Thus, 
$$ \partial f^{(n)} -n z^{n-1} = \partial {P}\bar{\partial}f^{(n)} 
= T\bar{\partial} f^{(n)}. $$
From where, 
$$\bar{\partial}f^{(n)} = \mu n z^{n-1} + \mu T\bar{\partial} f^{(n)}.$$

Suppose now $\|\mu\|_{\infty} \|T\|_{p} < 1 $. Then the solution of 
the last equation can be obtained by a Neumann series. More precisely,
 let $\phi_{1} = \mu n z^{n-1}$, and for 
$j>1$, $ \phi_{j} = \mu T \phi_{j-1} $. Hence,
$$ \|\phi_{j}\|_{p} 
\leq (\pi )^{1/p} \|t\|_{p}^{j-1} \|\mu\|_{\infty}^{j}.$$

\noindent  Hence $\sum _{i=1}^{\infty} \phi_{i}$ converges and  $
\bar{\partial} f^{(n)} =  \sum _{i=1}^{\infty} \phi_{i}$ and moreover
$\bar{\partial} f^{(n)} \in  L^p$. 

\medskip
This establishes the existence of the $f^{(n)}$. To construct a solution 
satisfying the initial condition 
$\int_{S^{1}}\omega^{(n)} = 0 $ we need the following:  

\begin{lm} \label{lemma:b2}			
Let $f$ be a quasiconformal mapping of the plane 
with complex dilatation $\mu$ of compact support, satisfying  
\[ \lim_{z \rightarrow \infty} f(z) - z^{n} - A_{0} = 0 \mbox{.} \]
 Then, 
$$ f(z) = z^{n} + A_{0} + {P}\sum_{i=1}^{\infty} \phi_{i} 
=  z^{n} + A_{0} + \sum_{i=1}^{\infty} {P}\phi_{i}.$$
\end{lm}

For a proof, see Theorem~4.3,  page 27 of \cite{lehto}. 
Although the proof there is for $n=1$ and $  A_{0}= 0 $, 
with minor modifications it gives the result we need. 
(Using the $\phi_{j} $ as we defined here.)

Hence, it it is easy to see, using Cauchy's theorem, 
that $\omega^ {(n)} = z^{n} + P\sum_{i=1}^{\infty} \phi_{i}(z) $,  
satisfies all the three conditions in (\ref{eq:5}). 
This concludes the proof of the theorem. 
\end{proof}

\subsection*{Acknowledgments}
J.~P.~Z. was supported by the Brazilian National Research Council
(CNPq, MA 521329/94-9). 
M.~S. and A.~T. were supported by the NSF. 
The conclusion of this work was made possible by the support of 
IMPA at the occasion of the 20th Brazilian Colloquium and the
4th PDE's Workshop.
A. T. would like to thank V. Arnold for explaining his point of view on the
Euler equation, as well as helpful conversations with D.~Sullivan, 
and
B.~Khesin for pointing out a number of important references. 
J.~P.~Z. would like to thank Welington de Melo for many interesting questions 
and conversations. The three 
authors would like to thank J.J.~Duistermaat, T.~Ratiu, and H.~Widom 
for stimulating discussions and helping with references.


\bibliographystyle{plain}

\end{document}